\documentclass[conference]{IEEEtran}
\usepackage{cite}

\ifCLASSINFOpdf
  \usepackage[pdftex]{graphicx}
\else
\fi
\usepackage{xurl}

\usepackage{float}
\usepackage{algorithm}
\usepackage{algpseudocode}

\usepackage{listings}
\usepackage{xcolor}

\usepackage{array}

\usepackage{booktabs}

\usepackage[hidelinks]{hyperref}

\usepackage{stfloats}

\definecolor{lstbg}{RGB}{248,248,248}

\algrenewcommand\alglinenumber[1]{\footnotesize #1:}
\usepackage{amsmath}
\begin{document}

\sloppy

\title{When Connectivity Is Not Enough: Cross-Layer Attacks on UAV C2 over 5G}

\author{\IEEEauthorblockN{Wagner Comin Sonaglio$^{*}$, Ágney Lopes Roth Ferraz$^{*}$, André Elias Melo$^{*}$, Murray Evangelista de Souza$^{*}$,\\
Sidnei Barbieri$^{*}$, Henrique Curi de Miranda$^{*}$, Guevara Noubir$^{\dagger}$ and Lourenço Alves Pereira Júnior$^{*}$}
\IEEEauthorblockA{$^{*}$Aeronautics Institute of Technology\\
Email: sonaglio@ita.br, roth@ita.br, andre.melo@ita.br, murraymes@ita.br, sidneisb@ita.br, henriquecuri1@proton.me, ljr@ita.br}
\IEEEauthorblockA{$^{\dagger}$Northeastern University\\
Email: g.noubir@northeastern.edu}
}


\maketitle

\begin{abstract}

Beyond Visual Line of Sight (BVLOS) unmanned aerial vehicle (UAV) operations increasingly use 5G standalone (SA) networks for command and control (C2) between the UAV and the ground control station (GCS). The 3rd Generation Partnership Project (3GPP) has specified mechanisms for authentication and authorization of unmanned aircraft systems (UAS) in this architectural setting. As a result, operators may treat registration state, Protocol Data Unit (PDU) session status, and IP reachability as evidence that the C2 path is available. In practice, however, these connectivity indicators alone do not guarantee that closed-loop control remains operationally safe. Attacks can degrade UAS C2 when timeliness degrades under shared User Plane contention, mobility continuity fails during Control Plane instability, or command integrity is violated at a trusted next-generation Node B (gNodeB). Such failures undermine connectivity as the central security indicator for UAV operations. In this paper, we demonstrate these issues using three distinct threat models on a reproducible Open5GS and UERANSIM testbed that carries Micro Air Vehicle Link (MAVLink) over the 5G User Plane, and we use a commercial Nokia core to ground deployment assumptions. We address timeliness, availability, and integrity through experiments in which attack success is defined as forcing an unsafe closed-loop state without a clean disconnect. We observe stale telemetry and heavy-tailed delay under co-tenant User Plane contention, failsafe after handover under Control Plane instability, and navigation hijacking after command rewriting at a compromised gNodeB. We further discuss why each threat model arises and evaluate mitigations for these cross-layer failures. Across the study, we disclosed five robustness issues, all with assigned CVEs.

\end{abstract}


\section{Introduction}
\label{sec:introduction}

BVLOS UAV operations are migrating from short-range radios to wide-area 5G SA transport between the ground control station (GCS) and the UAV \cite{baguer_enabling_2024,singh_overview_2022,geraci_what_2022}. This migration is attractive for such operations because 5G offers wide coverage, mobility support, and standardized integration for UAS services. 3GPP has already standardized support for UAS and C2 over 5G \cite{3gpp_ts_23256,singh_overview_2022}. However, the operational goal is not merely to connect a UAV to the network. The goal is to maintain safe closed-loop behavior in which commands and telemetry remain timely, continuous, and semantically trustworthy throughout the mission \cite{nassi_sok_2021,branco_cyber_2025}.

It is important to distinguish these terms because connectivity is only a prerequisite for control. A UE can remain registered, the PDU Session can stay established, and IP packets can still traverse the 5G User Plane while failures occur in the C2 loop. UAS authorization addresses association questions, such as which UAV and which controller are authorized to participate in the operation~\cite{3gpp_ts_23256,3gpp_ts_33256,3gpp_ts_22125}. Nevertheless, these mechanisms do not constrain shared User Plane contention, do not guarantee session continuity during Control Plane instability, and do not remove the trust breakpoint at the gNodeB after PDCP deciphering. As a result, indicators such as registration state, PDU Session state, and IP reachability can provide a misleading view of the UAV's operational security.

Existing 5G security studies often report results at the subscriber, protocol, or network core level, such as exposed interfaces, malformed-message failures, protocol deviations, or core function unavailability \cite{zhang_invade_2025,bennett_ransacked_2024,dong_corecrisis_2025,sun_5gc-fuzz_2025}. Existing UAV C2 studies frequently focus on the application layer, on MAVLink weaknesses, or on attacks against the radio link \cite{nassi_sok_2021,du_exploiting_2024}. These works are important but leave a gap at the boundary between mobile network security and cyber-physical control. In contrast, we ask a control-oriented question: what happens after the UE is already connected and C2 becomes ordinary IP traffic over the 5G User Plane?

In this work, we study the post-attachment gap within a unified UAV C2-over-5G SA experimental framework. We fix a readiness baseline that operators would normally consider healthy: the UE holds a valid access security context, the PDU Session is established, and MAVLink has bidirectional IP reachability. We then stress three interfaces that remain relevant after that baseline is met. TM1 stresses shared queueing on the User Plane. TM2 stresses service-based interfaces by sending black-box HTTP/2 requests to the Network Repository Function (NRF) and the Session Management Function (SMF) during mobility. TM3 stresses the radio access network (RAN) trust boundary by rewriting cleartext payloads on N3 after radio-layer deciphering on a compromised gNodeB.

Our observation is that attack success against UAV C2 should not be defined only as disconnection. In our experiments, the UE may remain registered, the data path may still exist, and the GTP-U tunnel may remain active, while the UAV receives stale telemetry, fails to recover during mobility, or executes a semantically altered command. The network still appears connected, but the control loop is no longer safe.

In summary, this paper makes the following contributions:

\begin{itemize}
    \item \textbf{Post-attachment threat models.}
    We define three UAV C2 threat models over 5G SA, covering User Plane contention, Control Plane instability during mobility, and RAN-side command tampering after PDCP deciphering.

    \item \textbf{5G SA UAV C2 testbed.}
    We build a reproducible Open5GS/UERANSIM testbed that carries MAVLink over the 5G User Plane and uses a Nokia CMU 5G SA core to ground the reachability and traffic-placement assumptions.

    \item \textbf{Connectivity is not safe control.}
    We show that C2 can suffer stale telemetry, \texttt{failsafe} during handover, or command hijacking while registration, IP reachability, or the GTP-U tunnel remain active.

    \item \textbf{Mitigation analysis.}
    We evaluate slice/DNN isolation, forwarding filtering, SBI segmentation with mTLS, and MAVLink signing, showing what each mitigation restores and what remains exposed.
\end{itemize}

Overall, our results show that connectivity-centric metrics can misread risk for cyber-physical C2 over 5G SA. Safe UAV operations require evaluating the control loop itself under adversarial contention, Control Plane instability, and RAN compromise, rather than treating registration and IP reachability as sufficient evidence of security.

The remainder of this paper is organized as follows. Section~\ref{sec:background} summarizes the minimum 5G and C2 concepts and presents the testbed. Section~\ref{sec:threat-models} presents the threat models and experiments. Section~\ref{sec:discussion} discusses security implications and limitations. Section~\ref{sec:related-works} positions the work relative to prior literature. Section~\ref{sec:conclusion} concludes, and Section~\ref{sec:ethics} discusses ethical considerations.

\section{Background}
\label{sec:background}

This section establishes the foundations for understanding the later sections, for example, how a PDU Session places C2 on the User Plane, where the SBI sits in the Control Plane, how N3 carries GTP-U, and where PDCP security terminates at the gNodeB. We also present basic concepts of UAVs and the MAVLink protocol, as well as the testbeds used in our experiments.

\subsection{Mobile Networks and 3GPP 5G}
\label{sec:background-3gpp}

In this work, we use the 5G SA architecture as the communication medium for BVLOS UAVs. In this architecture, radio-layer protection terminates at the gNodeB. The 5GC is IP-native and is anchored in SBA functions that interact via HTTP/2 on the SBI. Among these functions, the NRF supports discovery and profiles, the AMF handles registration and mobility, and the SMF manages PDU Sessions and interacts with the UPF on the N4 interface using PFCP~\cite{3gpp_ts_29244}. The UPF anchors user traffic and terminates GTP-U tunnels from the gNodeB toward the data network~\cite{3gpp_ts_23501,3gpp_ts_23502}.

\begin{figure}[H]
\centering
\setlength{\abovecaptionskip}{2pt}
\setlength{\belowcaptionskip}{2pt}
\includegraphics[trim=12 20 12 16,clip,width=\columnwidth]{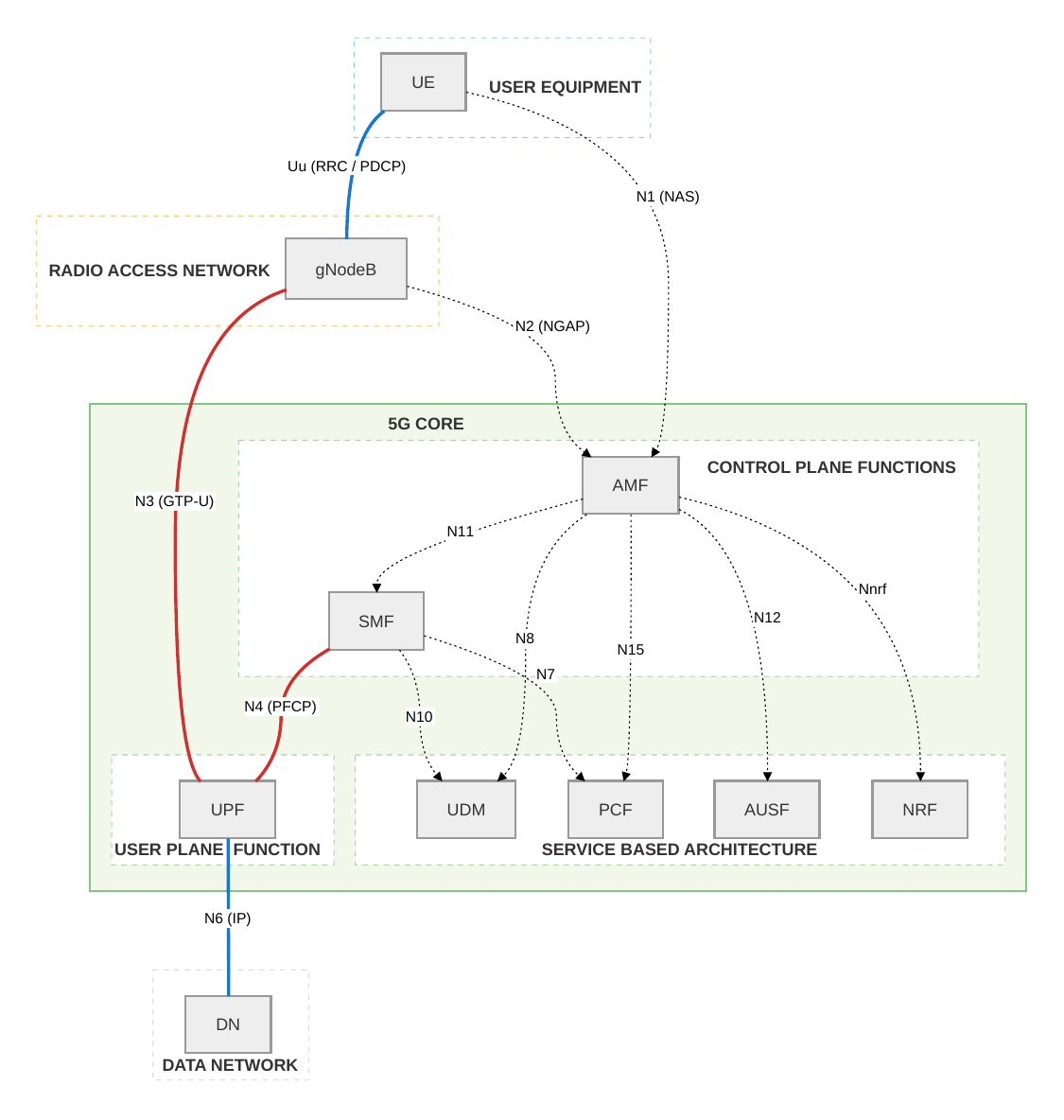}
\caption{Simplified 5G SA architecture and main interfaces.}
\label{fig:5g-architecture}
\end{figure}

Figure~\ref{fig:5g-architecture} summarizes the main elements of the 5G architecture. Traffic follows UE $\rightarrow$ gNodeB $\rightarrow$ UPF $\rightarrow$ data network within the PDU Session~\cite{3gpp_ts_23501,3gpp_ts_23502}. The S-NSSAI selects the logical slice, while the DNN selects the data network and the associated connectivity domain~\cite{3gpp_ts_23501,3gpp_ts_23502}. A detail we address later is that, within this configuration, two UEs can be placed in the same logical IP connectivity domain on the same slice and DNN if the operator does not impose additional isolation~\cite{3gpp_ts_23501}.

On the User Plane N3 interface, IP packets travel in GTP-U tunnels between the gNodeB and the UPF. Confidentiality and integrity on the radio interface apply only up to the gNodeB. PDCP protection terminates at the gNodeB~\cite{3gpp_ts_33501}. As a result, application bytes are in clear text in the RAN forwarding path before GTP-U encapsulation toward the UPF. User Plane integrity is optional in 3GPP and is often disabled in deployed 5G SA networks, mainly to avoid computational resource consumption~\cite{lasierra_fact-checking_2025,noauthor_bigmac_2023}. Without additional security layers, hop-by-hop trust prevails until the application adds end-to-end security to the communication.

\subsection{UAV Command and Control}
\label{sec:background-uav}

The GCS and the UAV exchange C2 over a bidirectional channel while the autopilot closes the control loop. Commands represent the operator's intent, and telemetry represents the observed aircraft state. Therefore, both must remain timely and consistent for the operation to be safe~\cite{nassi_sok_2021,branco_cyber_2025}. Loss, delay, or manipulation of C2 does not always appear as a complete link failure. These issues can appear as a gradual loss of control authority, an outdated state for the operator, or behavior divergent from what the pilot believes they are commanding. Thus, the vehicle may still respond to simple connectivity tests, such as \texttt{ping}, while no longer executing the expected operational intent.

\begin{figure}[H]
\centering
\includegraphics[width=\columnwidth]{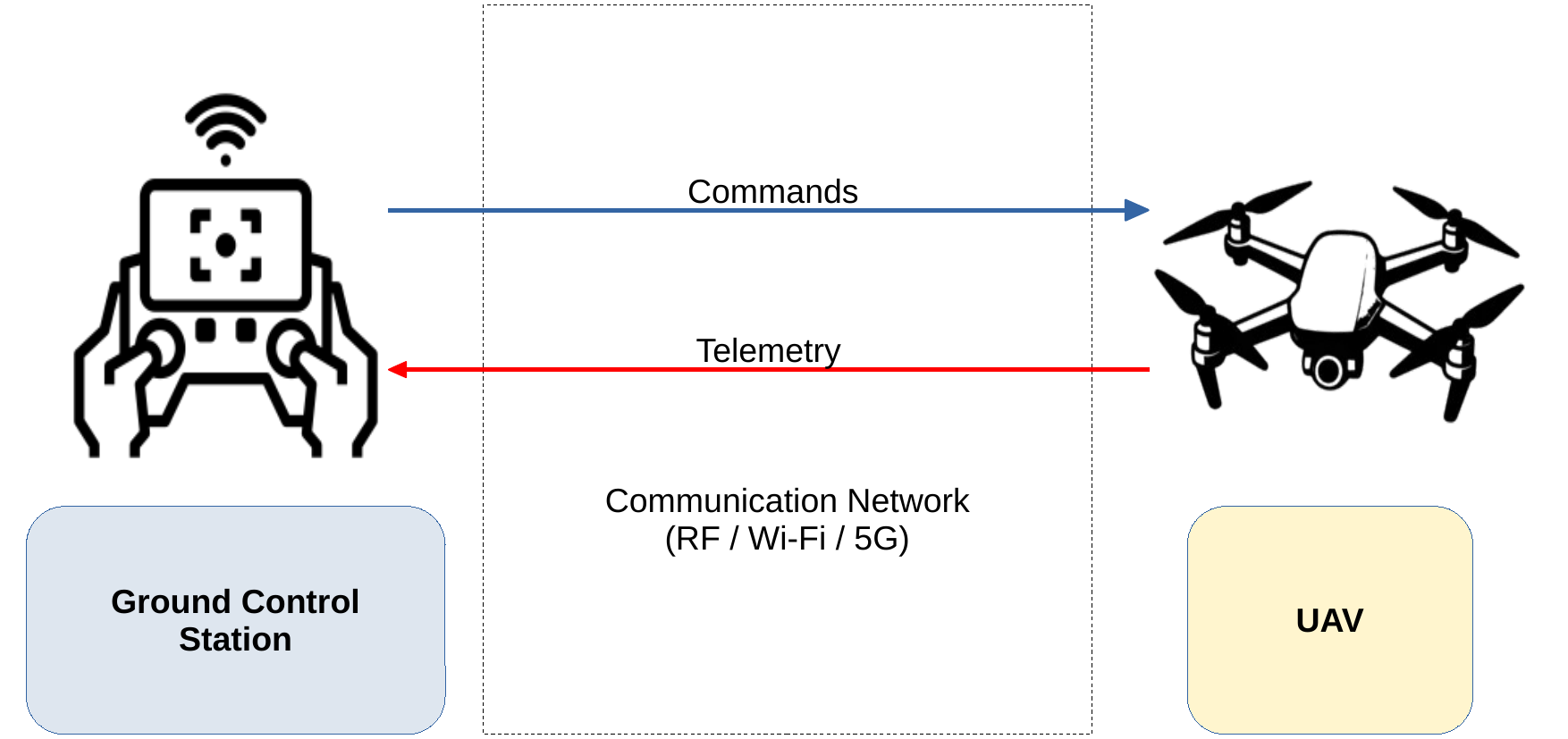}
\caption{Generic UAV command-and-control link over RF, Wi-Fi, or cellular networks.}
\label{fig:uav-commucation}
\end{figure}

Figure~\ref{fig:uav-commucation} situates UAV/GCS communication in a mobile network. Cellular networks, such as 5G, are relevant for BVLOS operations because they offer wide coverage, mobility support, and integration with long-range communication infrastructure~\cite{geraci_what_2022,baguer_enabling_2024,singh_overview_2022}. However, this type of operation does not depend solely on connectivity. Closed-loop control requires bounded delay, commands delivered within the operational window, and telemetry coherent with the actual UAV state. Therefore, in this work, we treat C2 as a cyber-physical workload, in which connectivity indicators can remain nominal even when the timeliness, continuity, or integrity of control has already failed.

On the 5G User Plane, MAVLink packets are carried over UDP~\cite{du_exploiting_2024,veksler_catch_2024,mavlink_guide,3gpp_ts_23501}. MAVLink messages carry commands and telemetry with semantic meaning for the receiver. In our setup, MAVLink serves as a representative payload for autopilot and GCS ecosystems and emulates the C2 loop over 5G IP. In this sense, our MAVLink workload acts as a lightweight digital twin of the UAV C2 loop: it preserves the command and telemetry semantics observed by the GCS and UAV while allowing repeatable network-level stress experiments over 5G IP. UDP favors low latency, but it does not provide reliable delivery guarantees. Checksums detect bit errors but do not authenticate the received content. Thus, an on-path adversary who observes cleartext packets can rewrite fields while preserving valid checksums.

MAVLink~2 signing closes this authenticity gap when enabled~\cite{ardupilot_mavlink2_signing}. It does not provide confidentiality. Some deployments keep signing disabled to simplify UAV/GCS integration~\cite{ardupilot_mavlink2_signing,px4_mavlink_encryption_2025}. This choice defines the baseline for TM3 before we enable signing as a mitigation.

In the 5G stack, C2 is ordinary IP traffic. Nonetheless, queues, session continuity, and trust boundaries below the application layer fall outside the scope of a simple connectivity test.

\subsection{5G UAV Experimental Testbed}
\label{sec:background-testbed}

To build a 3GPP-aligned testbed, we use Open5GS for the 5GC and UERANSIM for the gNodeB and UEs, tools employed in several prior studies~\cite{zhang_invade_2025,xu_integrity_2025}. The 5G stack runs on Kubernetes using Gradiant Helm charts on a Kind cluster on a single host~\cite{open5gs,kubernetes,open5gs_operator,kind,docker,ueransim}. After the PDU Session is established, each UE receives a \texttt{tun} interface. Python scripts using the MAVLink Python library~\cite{pymavlink} run in the UAV and GCS pods and exchange UDP packets over the path UE $\rightarrow$ gNodeB $\rightarrow$ UPF $\rightarrow$ peer UE. For the adversary, we instantiate different positions according to each threat model: a co-tenant UE in TM1, an adversary with SBI access in TM2, and a compromised gNodeB in TM3.

This software-defined environment lets us repeat runs and align logs while preserving the relevant 5G interfaces: PDU Sessions, GTP-U on N3, PFCP on N4, and SBI among core functions. The same slice, DNN, and MAVLink communication are used across TM1, TM2, and TM3, allowing us to isolate the adversary's position.

RF abstraction in the testbed is intentional. We do not report over-the-air latency. In exchange, we obtain deterministic stress on Control Plane and User Plane mechanisms on which BVLOS operations still depend in the field. We also carry MAVLink as a generic IP without instantiating all UAS NF procedures in this construction. TS~23.256 and TS~33.256 cover UAS authorization, UAS NF functions, USS, and controller-UAV pairing, including UUAA in TS~33.256~\cite{3gpp_ts_23256,3gpp_ts_33256,3gpp_ts_22125,3gpp_tr_33891}. Section~\ref{sec:discussion} revisits the contrast between these association mechanisms and the failure modes evaluated in this work.

Beyond the software testbed, we use the Nokia Compact Mobility Unit (CMU) as a commercial reference. We employ the CMU as a commercial 5G SA core to anchor deployment assumptions used by the threat models. The official documentation describes the CMU as a modular mobile packet core for private networks, with SA functions such as AMF, SMF, and UPF~\cite{nokia_cmu_2026}. In this environment, the UAV, GCS, and adversary hosts are connected behind Huawei AR502H-5G CPEs~\cite{huawei_ar502h-5g_2026} and appear as IP endpoints behind commercial 5G access. We use this configuration to ground reachability by slice and DNN, traffic placement, and endpoint behavior behind CPEs, not to exercise Open5GS-specific bugs.

Thus, Open5GS with UERANSIM remains the reproducible experimental substrate, while Nokia CMU with CPEs provides a commercial reality check for reachability and traffic behavior. In summary, TM1 connects shared slice and DNN to User Plane contention and C2 timeliness; TM2 connects the core and SBI to session continuity under mobility; and TM3 connects PDCP termination at the gNodeB to semantic integrity of MAVLink commands.

All code and data from this study are available on GitHub\footnote{\url{https://github.com/0wln3d/connectivity-not-enough-artifact}} to enable full reproducibility.

\section{Threat Models and Attacks}
\label{sec:threat-models}

To analyze the impact of cross-layer logical attacks on 5G networks and on the C2 link between the GCS and the UAV, we consider three scenarios that represent different adversary positions: (1) a malicious UE that is a co-tenant of the UAV and/or GCS, sharing the same slice and DNN, and able to reach the C2 link through the shared User Plane; (2) an insider adversary with access to the Control Plane, either through a UE that can improperly reach 5GC functions or through someone with direct access to SBA and SBI functions; and (3) a compromised gNodeB, reached through UE-accessible infrastructure, lateral movement, physical compromise, or someone with direct access to the Control Plane, in which the adversary has full control of the RAN node.

For each threat model, we describe the attacker's capabilities, assumptions, and assets at risk, along with an example attack in our test environments.

\subsection{Threat Model 1: Co-tenant Malicious UE (Same Slice and DNN)}
\label{sec:threat-rogue-ue}

TM1 represents a threat model focused on the degradation of C2 timeliness. The adversary in this scenario is a co-tenant UE assigned to the same slice and DNN as the UAV and/or the GCS, considering that in 5G networks, slices can be configured for services with stringent latency requirements, such as URLLC \cite{3gpp_ts_23501,3gpp_ts_22261}. This scenario is possible, given that the adversary can be a legitimate, compromised, or maliciously provisioned UE within the same slice/DNN. Previous studies show that UE/UAV impersonation can also lead to this positioning \cite{lisowski_simurai_2024}, but our attack does not depend on that condition. Furthermore, security misconfigurations, as observed in our tests on the Nokia core, may allow direct connectivity between UEs on the same data network/subnet.

In this scenario, the attacks do not depend on exploiting bugs in the 5GC, breaking authentication mechanisms, or modifying MAVLink commands. The attacks arise when UEs share the User Plane, enabling the adversary to generate traffic that interferes with C2 traffic for UAS operations.

The choice of slice, DNN binding, and the operator’s forwarding policy determines whether authorized UEs land in the same logical IP environment and share the same path through the UPF \cite{3gpp_ts_23501,3gpp_ts_23502}. Having more than one UE on the same slice and the same DNN does not, by itself, imply that the neighborhood is insecure. However, this configuration creates a condition in which User Plane isolation must be explicitly enforced to ensure security filtering between subscribers. Without this complementary isolation, as observed in our testbed and in tests on the Nokia core, C2 and adversarial traffic may contend for the same paths.

Prior work notes that slice association can expose privacy-relevant group information~\cite{olimid_5g_2020}. In addition, TS~23.501 defines VN groups for 5G LAN-type services, with unicast forwarding among members on the User Plane~\cite{3gpp_ts_23501}, a setting also relevant to UAS support~\cite{3gpp_ts_23256,3gpp_ts_23501}. We do not instantiate all VN Group procedures, but UE-to-UE reachability is a foreseen architectural design, not an emulator artifact.

TS~23.256 and TS~33.256 describe UUAA, UAS NF functions, authorization of the USS service provider for the UAV, and controller-UAV pairing \cite{3gpp_ts_23256,3gpp_ts_33256,3gpp_ts_22125,3gpp_tr_33891}. These mechanisms deal mainly with authentication, authorization, and pairing, that is, with who may participate in the UAS operation and which controller can associate with the UAV. TM1 is orthogonal to this scope because, even when the GCS is authorized, the UAV is authorized, and the C2 flow is permitted by the network, packet delivery can still be delayed enough to compromise the safe closure of the control loop \cite{singh_overview_2022,baguer_enabling_2024}.

\begin{figure}[H]
\centering
\setlength{\abovecaptionskip}{2pt}
\setlength{\belowcaptionskip}{2pt}
\includegraphics[width=\columnwidth]{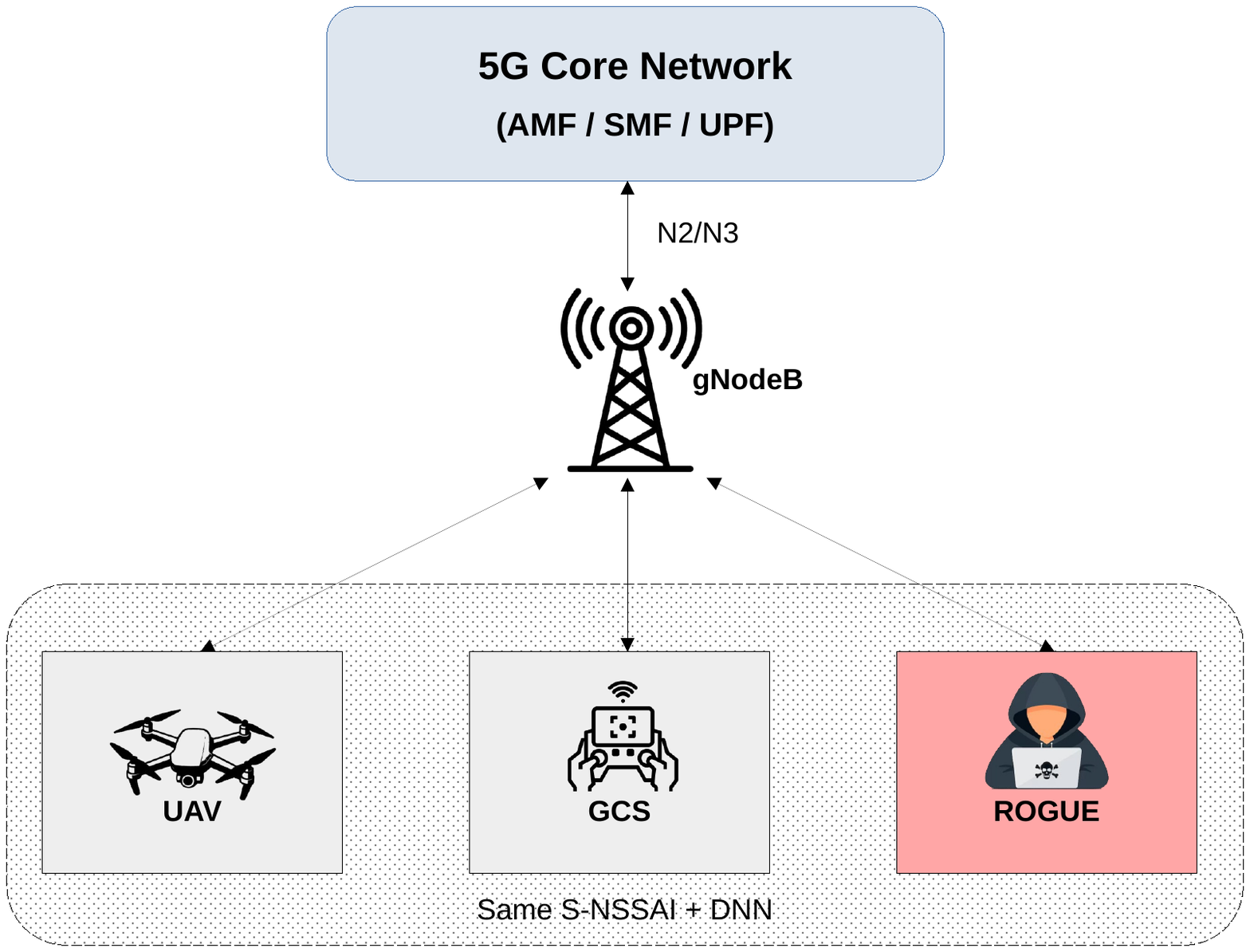}
\caption{TM1 logical placement of UAV, GCS, and ROGUE UE in a shared S-NSSAI and DNN.}
\label{fig:threat-model-01}
\end{figure}

Before we present the findings on traffic attacks, we use the configuration found in the Nokia core as ground truth. In the default configuration, the three UEs registered for the tests were on the same slice and DNN. After attaching, we identified bidirectional IPv4 reachability among these UEs without an overlay tunnel. This scenario matches what we observe in our testbed and the logical placement in Figure~\ref{fig:threat-model-01}, which shows that the co-tenant on a shared slice/DNN is an observable foothold in commercial core equipment, not merely a laboratory convention.

Figure~\ref{fig:threat-model-01} shows the ROGUE UE as a co-tenant of the UAV and GCS in the same slice and DNN, sharing User Plane forwarding resources up to the UPF.

After identifying that UE-to-UE reachability is possible, including basic IPv4 network scanning and conventional IPv4 traffic-stress attacks, we ran experiments on the Nokia core with two traffic variants. The first variant, UExUE, consists of the ROGUE UE sending sustained UDP datagram stress to the UDP port used for GCS-UAV communication (UDP Flood). In the second variant, we considered how traffic from one UE could affect another UE, even when direct same-slice/DNN placement is unavailable. We call this variant “boomerang”: in our observed configuration, a device connected to the same VLAN as the core could reflect traffic toward UE addresses reachable through the Nokia core. In the “boomerang” attack, which is a reflected DoS, a low-duty-cycle port-churn burst generator script sends short, spaced UDP bursts, rotating the destination port within a window, against the reflector’s IP.

Additionally, during our experiments, we identified and disclosed Open5GS CVE-2026-8187~\cite{mitre_cve-2026-8187_2026}, in which a UE with direct reachability to the UPF GTP-U endpoint may stress the UPF through the \texttt{\_gtpv1\_u\_recv\_cb()} function and cause C2 and RTT degradation similar to the other attacks (approximately 4$\times$ RTT increase and approximately 14.3\% MAVLink command loss). The issue was reported to the maintainer, and the CVE record is now public. This suggests that, given the possibility observed in the Nokia core that a UE may directly reach core-side infrastructure, attacks of this kind may represent a broader class of traffic-based timeliness and availability problems.

During preliminary calibration, we identified an operational sweet spot between minimal RTT-probe loss and observable MAVLink C2 degradation, using the parameters below:

\begin{table}[H]
\centering
\caption{TM1 traffic profiles used for attack calibration.}
\label{tab:tm1-attack-profiles}
\footnotesize
\setlength{\tabcolsep}{4pt}
\begin{tabular*}{\columnwidth}{@{\extracolsep{\fill}}lp{0.72\columnwidth}@{}}
\toprule
Scenario & Profile \\
\midrule
Nokia UE-to-UE & Constant UDP, 25~pps, 32 B payload \\
Nokia reflected DoS & Random-port bursts, 5~ms ON / 2500~ms OFF,\newline 800~B payload \\
Open5GS UE-to-UE & Constant UDP, 100~pps, 32 B payload \\
\bottomrule
\end{tabular*}
\end{table}

While conducting these attack experiments, we measured the end-to-end effect on C2. In tests on the Nokia core, we ran five experiments, each with baseline/attack/recovery phases lasting 3 minutes. We captured RTT, \texttt{timeout} events, delays in delivered commands, and commands issued by the GCS but not received by the UAV within the log window. Table~\ref{tab:tm1-command-delivery} and Table~\ref{tab:tm1-rtt-summary} summarize the collected data. Raw CSVs and command traces are included in the artifact for line-by-line auditing.

\begin{table}[H]
\centering
\caption{C2 command delivery under TM1 traffic variants.}
\label{tab:tm1-command-delivery}
\footnotesize
\setlength{\tabcolsep}{4pt}
\begin{tabular*}{\columnwidth}{@{\extracolsep{\fill}}lrrrr@{}}
\toprule
Scenario & Sent & Received & Lost & Loss (\%) \\
\midrule
Nokia UE-to-UE & 210 & 198 & 12 & 5.71 \\
Open5GS UE-to-UE & 126 & 106 & 20 & 15.87 \\
Nokia Reflected DoS & 209 & 175 & 34 & 16.27 \\
Open5GS Reflected DoS & 126 & 101 & 25 & 19.84 \\
\bottomrule
\end{tabular*}
\end{table}

\begin{table}[H]
\centering
\caption{RTT summary for TM1 traffic variants, in milliseconds.}
\label{tab:tm1-rtt-summary}
\footnotesize
\setlength{\tabcolsep}{3pt}
\begin{tabular}{@{}lrrrrrr@{}}
\toprule
Scenario & Probes & Loss (\%) & Median & p95 & p99 & Max \\
\midrule
Nokia UE-to-UE          & 25,333 & 2.0 & 24.63 & 31.28 & 600.33 & 958.90 \\
Nokia Reflected DoS    & 17,258 & 2.8 & 67.25 & 90.38 & 618.03 & 995.38 \\
Open5GS UE-to-UE       & 31,081 & 0.1 & 0.98  & 1.80  & 3.30   & 101.65 \\
Open5GS Reflected DoS  & 31,013 & 0.1 & 1.02  & 2.00  & 4.03   & 18.50  \\
\bottomrule
\end{tabular}
\end{table}

\begin{figure}[H]
\centering
\includegraphics[width=\columnwidth]{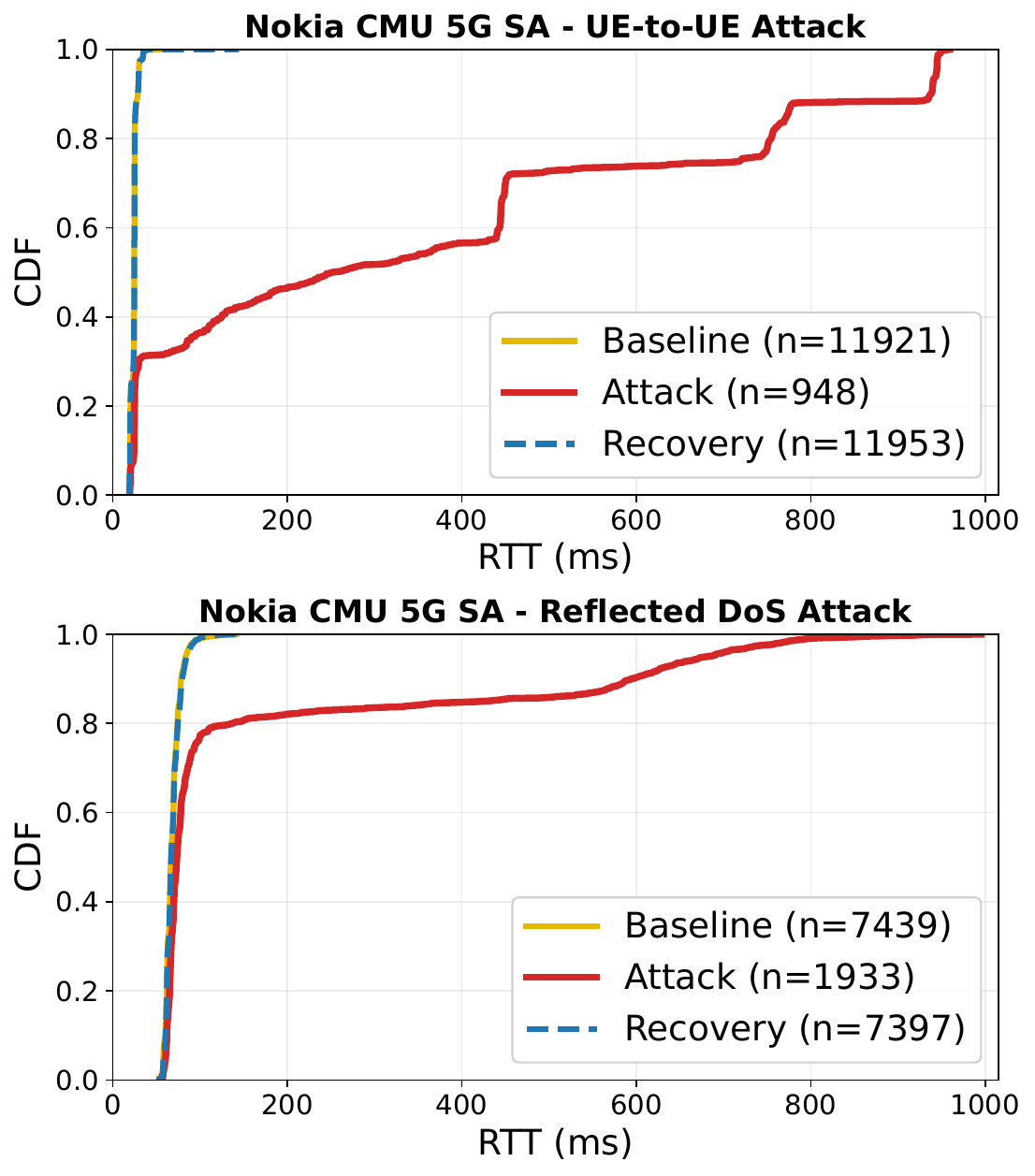}
\caption{RTT CDFs for TM1 traffic variants on the Nokia core.}
\label{fig:threat-model-01-cdf}
\end{figure}

Figure~\ref{fig:threat-model-01-cdf} shows, for each attack variant on the Nokia core, the empirical CDF of RTT computed only over completed probes, with baseline, attack, and recovery phases. The baseline and recovery phases overlap almost completely and remain concentrated at low RTTs, thereby confining the worst behavior to the attack interval. The two attack variants affect C2 in different ways. In the UExUE attack (UDP flood), the attack curve spans a wide RTT range and shows large steps, reaching almost 1 second. In the reflected DoS, most completed probes still remain near the baseline, but the distribution develops a long tail. This suggests a more intermittent effect, in which the attack need not keep the path continuously saturated to disrupt C2. Short contention windows suffice for MAVLink commands to be missed or delivered too late within the control window. This behavior is consistent with the reflected variant stressing intermediate forwarding or CPE/NAT processing via random-port bursts, although we treat this as an operational explanation not a claim about the internal implementation of commercial equipment.

\begin{figure}[H]
\centering
\includegraphics[width=\columnwidth]{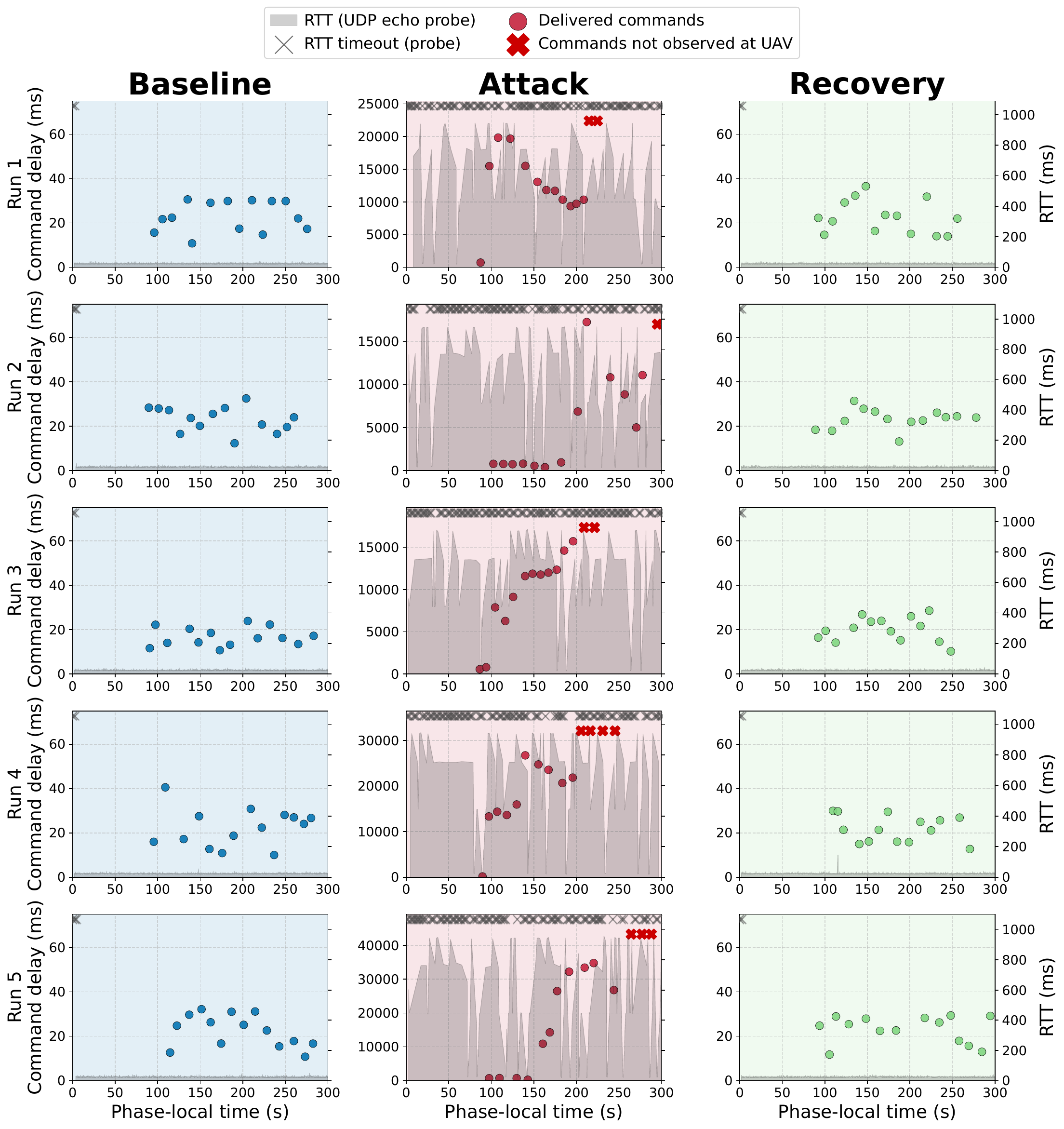}
\caption{Per-run TM1 timeline for the Nokia UE-to-UE attack.}
\label{fig:threat-model-01-nokia-timeline}
\end{figure}

Figure~\ref{fig:threat-model-01-nokia-timeline} complements the aggregated CDF by showing the temporal behavior of each Nokia UExUE run. The figure shows that baseline and recovery remain stable, with low command delay and completed RTT probes near the normal range. During the attack, however, RTT probes repeatedly time out, and several MAVLink commands are delayed or not recorded in the UAV's log window. 

One interesting detail we observed is that in the UExUE attack, the UDP flood traverses the 5G User Plane path toward the target UE, thereby directly contending for the same queues and forwarding resources used by RTT probes and C2 traffic, increasing RTT timeouts and missed MAVLink commands. Nevertheless, in the reflected DoS, or “boomerang”, the bursts, although initiated from a UE, reach the target UE from a core-side segment, reducing persistent occupancy of the most critical queues and producing a more intermittent effect on C2, with lower RTT-probe loss. This suggests that the traffic injection point changes the failure shape, but not the security implication, since even lower RTT-probe loss can still coexist with delayed or missed MAVLink commands.

Overall, this exposes a gap between 5G connectivity and safe C2 controllability. Together, the CDF and timeline views show that, in the TM1 scenario, a complete disconnection is not necessary to affect C2. The UAV and the GCS remain connected, and the path remains partially usable. Even so, brief contention periods suffice to make the control loop unreliable. Thus, the failure is not captured by median RTT alone, but by the combination of RTT timeouts, delayed commands, and commands not observed by the UAV during the attack interval.

The previous results focused on the Nokia core. Nevertheless, as Table~\ref{tab:tm1-command-delivery} and Table~\ref{tab:tm1-rtt-summary} show, the RTT measured in our testbed experiments was lower because the User Plane path is almost entirely software co-located in pods over a virtual network, using Kind, container bridges, the same host, or only a few lab hops. Unlike the Nokia CMU 5G SA setup, our testbed does not include a real radio link, external transport segments, or the additional bottlenecks present in a commercial 5G SA deployment. Even so, as on Nokia, C2 degradation followed the same pattern, supporting the observation that connectivity does not guarantee safe UAS C2.

To mitigate these issues in TM1, we evaluated mitigations only in our Open5GS testbed with UERANSIM. Although evaluated only in this testbed, these mitigations are not specific to Open5GS. They rely on deployment-level controls, namely slice/DNN separation and forwarding filtering, and should therefore be portable to 5G SA deployments that expose equivalent isolation and filtering mechanisms. Their exact effectiveness, however, depends on operator configuration and on where these controls are enforced.

For the UExUE scenario, we placed the ROGUE UE on a different S-NSSAI and DNN than those of the UAV and GCS, removing Layer 3 adjacency. For the reflected DoS attack, we used \texttt{iptables} filtering at the forwarding boundary to drop reflected UDP traffic toward the ports used by the UAV and GCS. We examined overhead in both cases and observed no measurable increase in RTT or MAVLink command delay. These results indicate that the mitigations are practical, but their effectiveness depends on correct operator configuration.

Finally, TM1 documents a timeliness failure. We show that the network can remain registered and reachable while MAVLink C2 becomes operationally unsafe relative to actuation deadlines \cite{singh_overview_2022,baguer_enabling_2024}. Section~\ref{sec:discussion} revisits UUAA and pairing for readers who seek contrast with UAS association policies foreseen by 3GPP \cite{3gpp_ts_23256,3gpp_ts_33256}.

\subsection{Threat Model 2: Insider Control-Plane Adversary (SBI Reachability)}

For TM2, we address a scenario in which the attack targets Control Plane availability and its impact on UAS C2. Because BVLOS UAVs rely on session continuity for long-distance operations, 5GC coordination is fundamental to maintaining the IP session that carries MAVLink traffic.

The adversary does not need the same slice/DNN as the UAV, or access to its User Plane path. It needs IP reachability into the operator core so that HTTP/2 SBI endpoints of functions such as the NRF or SMF can be reached. Misconfigured routing and segmentation can push traffic across trust boundaries into sensitive internals~\cite{shaik_uncovering_2025}. Insiders can reach the same vantage point through compromised orchestration, abused administration, supply-chain compromise, or weak separation between management and service planes~\cite{shaik_uncovering_2025}. Penetration tests on widely used open-source 5G cores show that these HTTP/2 interfaces carry familiar web risks~\cite{giambartolomei_penetration_2024}. TM2 is about SBI exposure once perimeter assumptions fail, not about sharing the UAV's User Plane path.

In 5G service-based architectures, NFs expose APIs over HTTP/2. Schema checks may pass, yet requests may still be inconsistent with an NF's internal state. This mismatch is the technical axis of TM2. Schema validity does not guarantee semantic validity with respect to the NF runtime state. This makes TM2 representative of a broader class of 5GC failures: requests that are syntactically valid at the SBI boundary but semantically inconsistent with NF state can propagate from Control Plane instability to User Plane session continuity.

To anchor plausibility while avoiding claims about vendor-specific bugs, we use two ground references. First, during our experiments in the Nokia core lab, we verified that the core SBA functions were protected within a Layer 3 Virtual Routing and Forwarding (VRF) domain. Nonetheless, when running \texttt{traceroute} from the UE toward NF/SBI address ranges and toward the gNodeB, we observed intermediate IP nodes in the core VLAN. Scanning these IPs revealed exposed SSH ports and web services on transit devices. A compromise of such assets could therefore provide a pivot toward functions that were otherwise isolated from the UE data plane. Thus, the Nokia core is used here as operational ground truth for reachability and segmentation assumptions, not as evidence of the specific Open5GS vulnerabilities. Second, Open5GS provides a reproducible SBA instance to exercise a concrete propagation chain. Therefore, as in other works \cite{sun_5gc-fuzz_2025}, we use Open5GS to adversarially explore these possibilities in a 3GPP-aligned network and observe the impact on C2.

\begin{figure}[H]
\centering
\includegraphics[width=\columnwidth]{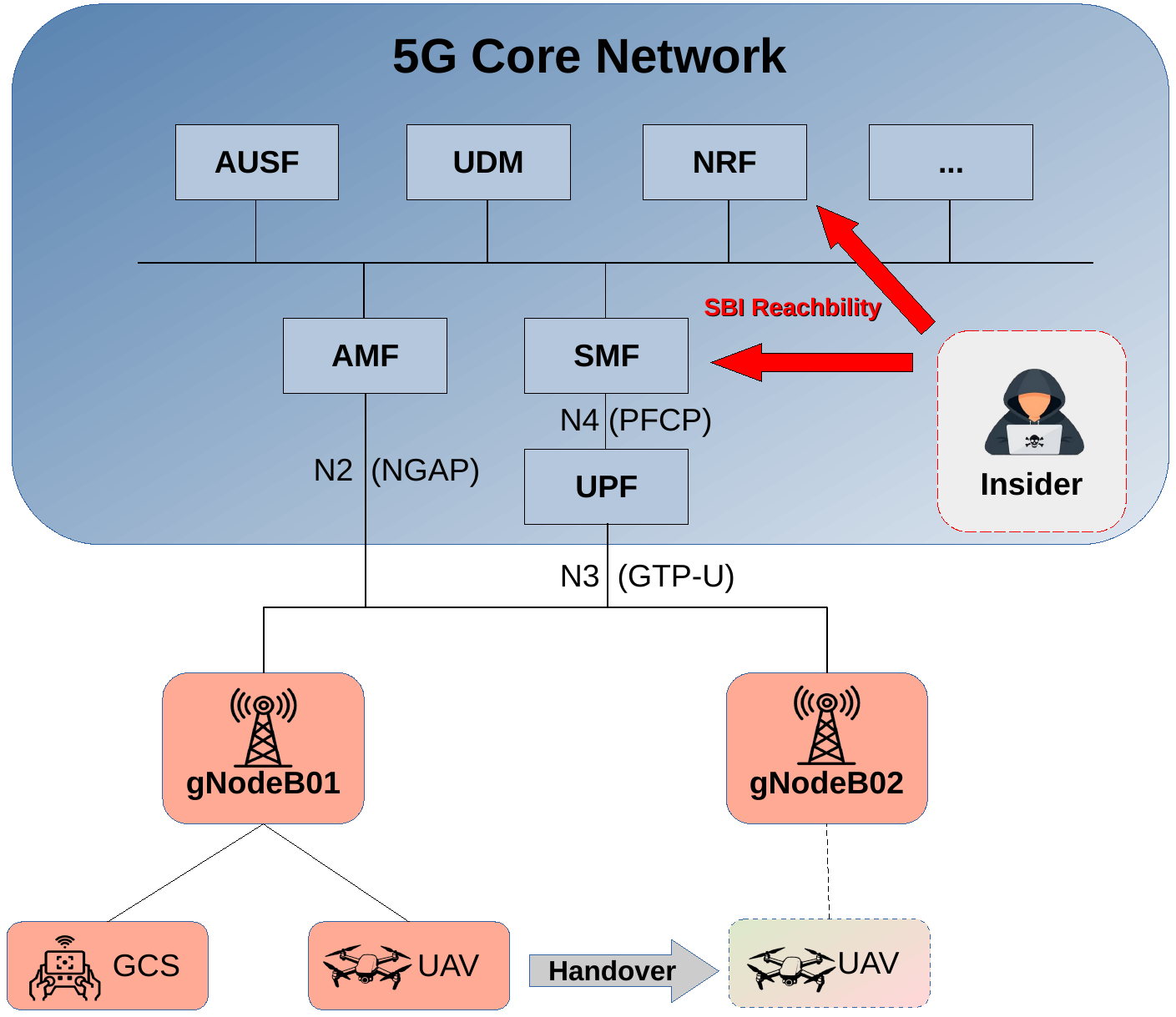}
\caption{TM2 Control Plane attack surface during UAV handover.}
\label{fig:threat-model-02}
\end{figure}

As with TM1, Open5GS runs the NFs in Kubernetes. UERANSIM provides gNodeBs and UEs. The UAV and the GCS exchange C2 traffic over UDP on the User Plane. The difference is that TM2 uses two gNodeBs to exercise handover, which approximates the mobility component of BVLOS operation, as shown in Figure~\ref{fig:threat-model-02}. For the adversary, which we call an “insider”, we instantiate a pod that can reach the SBI and issue HTTP/2 requests to SBA-exposed NFs, such as the NRF and SMF.

\begingroup
\setlength{\dbltextfloatsep}{6pt plus 2pt minus 2pt}
\begin{figure*}[!t]
\centering
\setlength{\abovecaptionskip}{2pt}
\setlength{\belowcaptionskip}{2pt}
\includegraphics[trim=0 14 0 10,clip,width=\textwidth]{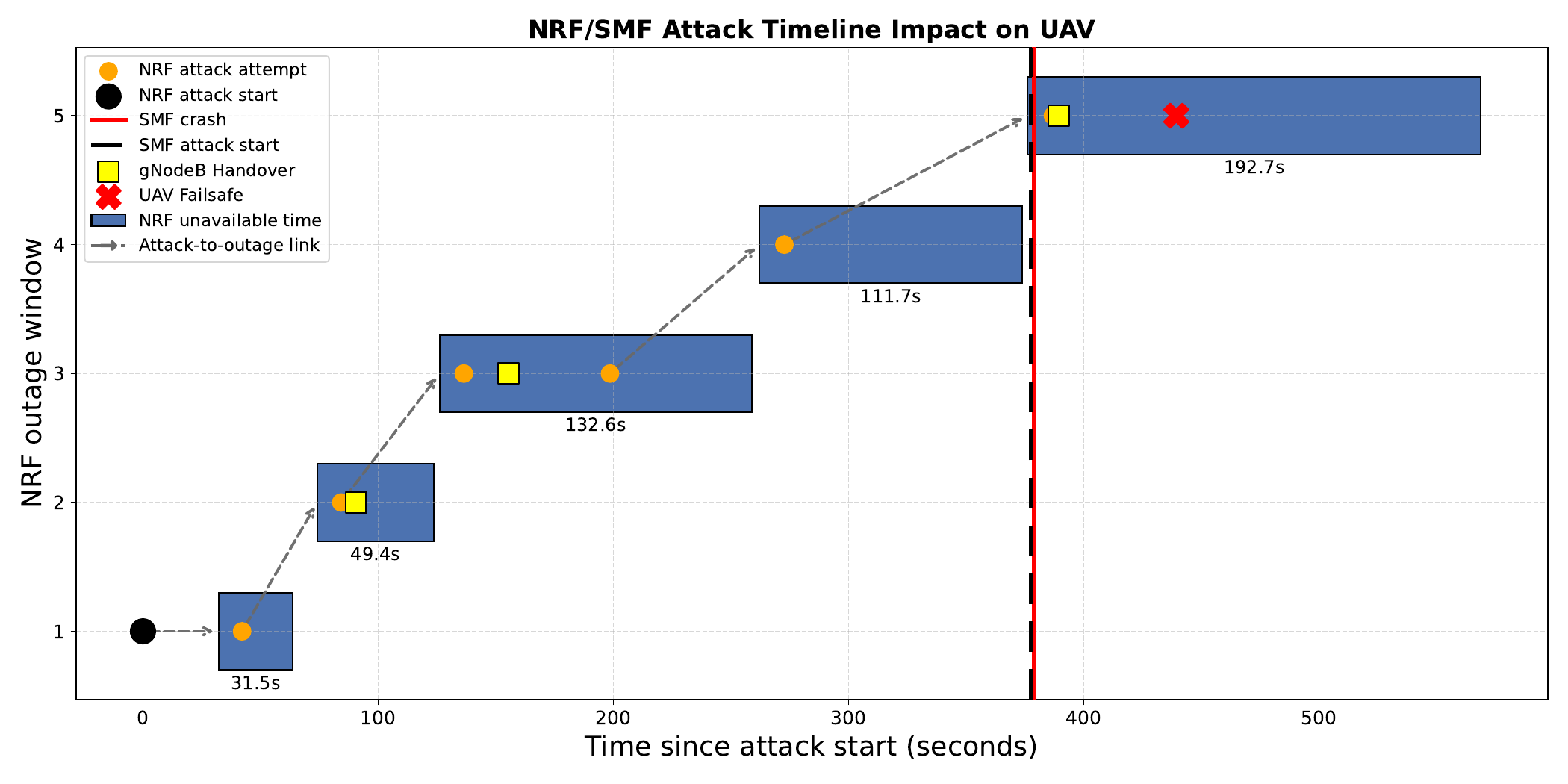}
\caption{TM2 timeline of NRF outage, SMF crash, handover, and UAV failsafe.}
\label{fig:threat-model-02-attack-timeline}
\end{figure*}
\endgroup

While exploring this scenario in Open5GS, we found, in addition to the UPF degradation bug discussed in TM1, three additional robustness issues. The UPF crash under crafted GTP-U traffic is tracked as CVE-2026-8186~\cite{mitre_cve-2026-8186_2026}, and the SBI issues affecting NRF and SMF are tracked as CVE-2026-38511~\cite{mitre_cve-2026-38511_2026} and CVE-2026-38512~\cite{mitre_cve-2026-38512_2026}. The Open5GS maintainer was contacted through the corresponding issue reports and has already addressed the reported failures.

In the first bug (CVE-2026-38511)~\cite{mitre_cve-2026-38511_2026}, a DoS is possible from an insider against the Open5GS NRF through a reachable assertion in the SBI notification-emission path. Specifically, a \texttt{nfStatusNotificationUri} field in \texttt{http://host:port} form, without a path component, causes \texttt{ogs\_sbi\_getpath\_from\_uri()} not to populate \texttt{path}, leading to an \texttt{ogs\_assert()} and a crash of \texttt{open5gs-nrfd}. In the second bug (CVE-2026-38512)~\cite{mitre_cve-2026-38512_2026}, an insider with a valid \texttt{pduSessionRef} can crash \texttt{open5gs-smfd} via a crafted \texttt{POST} request to \path{/nsmf-pdusession/v1/pdu-sessions/{pduSessionRef}/modify}. The request carries \texttt{HsmfUpdateData} that combines \texttt{requestIndication = UE\_REQ\_PDU\_SES\_MOD} with \texttt{upCnxState = SUSPENDED}, creating a mismatch with the internal session state and \texttt{pfcp\_flags}, and driving \texttt{smf\_gsm\_state\_operational()} into a fatal invalid-state failure. In the third bug (CVE-2026-8186)~\cite{mitre_cve-2026-8186_2026}, the UPF GTP-U surface on UDP 2152 can terminate with \texttt{SIGSEGV} under sustained remote traffic with crafted GTP-U datagrams. In the reported setup, this occurred on the order of tens of seconds and required an active-session TEID to reach a deeper packet-processing path, where insufficient validation during IPv4 parsing can cause subsequent transport-header reads to access invalid memory. This causes User Plane DoS and leaves PDU Sessions unavailable until the UPF service is restarted. In the TM2 experiments, we use the bugs found in the NRF and SMF, and exercise them in two stages:

\textbf{Step 1: NRF instability loop.} Here we ran an exploit that keeps the NRF in a crash loop (\texttt{CrashLoopBackOff}), using the issue reported previously. In Kubernetes, successive restarts enter the \texttt{CrashLoopBackOff} state. The kubelet applies exponential backoff between restart attempts \cite{kubernetes_pod_lifecycle_2026}. As a result, NRF unavailability grows across cycles. In our runs, it reached approximately 395~s.

\textbf{Step 2: SMF crash during NRF unavailability.} After starting a \texttt{CrashLoopBackOff} with the NRF down for a considerable time, we ran a script that triggers the SMF bug reported previously. With the NRF in \texttt{CrashLoopBackOff}, after a restart, the SMF could not re-register with the NRF, remained in an inconsistent operational state, and new UE PDU Sessions were not possible.

To verify the impact of this inconsistent core state on the UAV mission, we triggered UAV mobility through a gNodeB handover. On a healthy core, there is a short interruption, and the User Plane returns. Under NRF and SMF instability, session continuity fails during handover, as we observed in our experiment. Mobility requires timely Control Plane coordination. The UAV UE does not reliably restore data connectivity. The GCS and UAV stop exchanging MAVLink heartbeats over UDP for an extended period. As a result, the UAV enters \texttt{failsafe} and aborts its mission. Figure~\ref{fig:threat-model-02-attack-timeline} summarizes the full TM2 experiment.

With these ground truths and experiments, TM2 shows that core availability is not merely a “network down” problem; it becomes an operational security problem for the UAS mission. Even when the attack does not touch the UAV, does not inject commands, and does not directly interfere with MAVLink traffic, a Control Plane failure can break session continuity at the moment when mobility requires coordination. Therefore, the UAV may remain “connected” up to a point, yet lose the ability to maintain safe navigation when the system must react to handover, session recovery, or state updates.

To evaluate mitigations following a 3GPP-aligned pattern, we reduced SBI reachability and hardened request handling. On the network plane, we used \texttt{NetworkPolicy} to limit access to the NRF and SMF to pods that implement legitimate NFs. Generalizing this to 5G deployments, including commercial offerings, ACLs and firewalls play an equivalent mitigation role in non-Kubernetes environments. On the transport and application planes, we used mTLS with per-NF identity to reduce the risk of unauthenticated reachability to SBI services. However, mTLS does not replace semantic request validation inside the NF. For completeness, we also tested transport-layer protections such as IPSec, which did not mitigate these application-layer failure modes in our setting. 

\begin{figure}[H]
\centering
\includegraphics[width=\columnwidth]{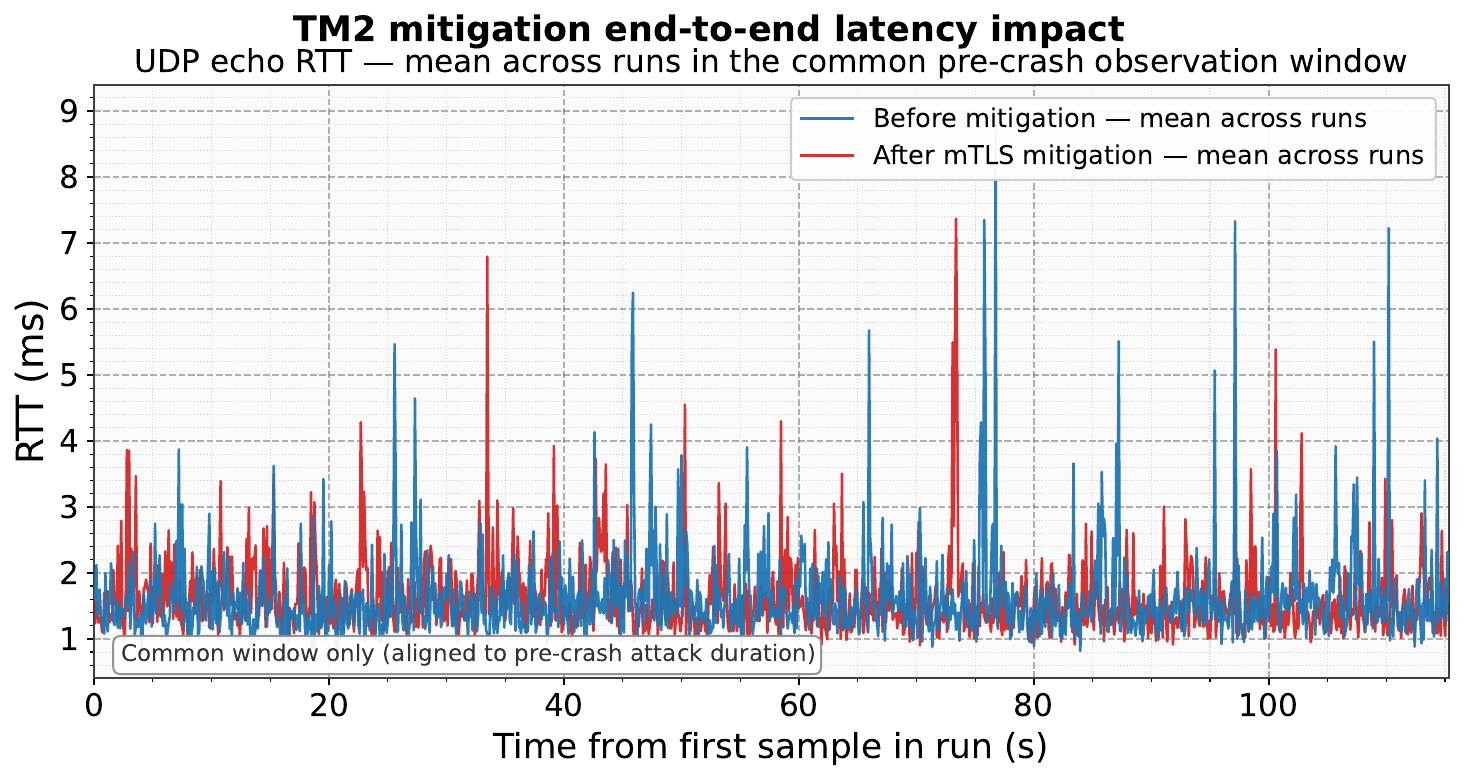}
\caption{TM2 C2 latency before and after mTLS mitigation.}
\label{fig:threat-model-02-mtls-overhead}
\end{figure}

This is relevant because operators may avoid or selectively deploy optional cryptographic mechanisms in the User Plane/Control Plane (such as IPSec) \cite{shaik_uncovering_2025}, particularly due to their operational and performance overhead \cite{noauthor_bigmac_2023}. With that in mind, we measured the impact of the mTLS mitigation on the UEs. We performed three runs for each condition (with and without mitigation) and compared C2 RTT across comparable windows during which the UAV remained available. Figure~\ref{fig:threat-model-02-mtls-overhead} compares the mean RTT of C2 traffic before and after mitigation with mTLS, using only the common pre-crash window when the UAV was still available. The two curves remain almost fully overlaid, with typical RTT around 1 to 2~ms and only occasional spikes under both conditions. This indicates that the mitigation reduced improper SBI reachability without introducing perceptible overhead for the UEs or degrading the MAVLink channel during the observed window.

TM2 documents an availability failure. Our experiments show that the network may appear connected before mobility, while C2 over MAVLink becomes operationally unsafe when Control Plane instability prevents session recovery during handover. Section~\ref{sec:discussion} revisits this cross-layer dependency, including Kubernetes-based deployments and cloud-native 5GC.

\subsection{Threat Model 3: Compromised RAN Node (C2 Command Tampering)}
\label{sec:threat-gNodeB}

To close out the trio of post-attachment logical failures we discuss in the paper, we address TM3, which shifts the focus to the semantic integrity of the C2 command stream. The path remains up, the flow continues, and even so, the UAV executes a different intent.

Research in the literature on the gNodeB tends to focus on fake gNodeBs and unauthorized RAN nodes \cite{xu_integrity_2025,shaik_uncovering_2025,luo_sni5gect_2025}. Our focus is different. Here, the gNodeB remains legitimate from the core's perspective. What changes is control of the forwarding host. The actor operates after PDCP deciphering, so the effect does not depend on breaking radio keys or a MAVLink parser bug. TR~33.809 discusses hardening against false base stations, including cell authenticity and network-side detection \cite{3gpp_tr_33809}.

Recent work explores threats and defenses centered on radio-interface attacks, integrity violations involving base stations, and failures of routing/isolation across deployment boundaries \cite{xu_integrity_2025,shaik_uncovering_2025,luo_sni5gect_2025}. These efforts do not fully cover the case in which a legitimately deployed gNodeB later falls under adversary control while still appearing as an authorized network element. We do not claim that compromising a gNodeB is trivial. Such compromise can arise from operational realities that expand the RAN attack surface and interoperability footprint, including disaggregated multi-vendor Open RAN deployments with open interfaces and third-party integration paths \cite{baguer_enabling_2024}, and from architectural assumptions about trusted internal interfaces and weak boundary enforcement that become dangerous when trust anchors shift at deployment time \cite{shaik_uncovering_2025}, as well as from mismatches between security expectations and what vendors and operators actually deploy in commercial networks \cite{lasierra_fact-checking_2025}.

TM3 should therefore not be read as a claim that gNodeB compromise is easy, nor that a compromised forwarding node rewriting cleartext traffic is surprising in isolation. It isolates the consequences of that compromise for UAS C2: connectivity, radio-layer protection, and GCS-side session health can all remain normal even as the autopilot receives a semantically different command.

As in TM2, we use the Nokia core as operational ground truth for reachability and segmentation assumptions, not as evidence of commercial gNodeB compromise. Traceroute measurements from the UE exposed intermediate IP nodes along the path toward the RAN and core-side infrastructure, showing that the forwarding elements supporting the User Plane are observable operational elements rather than an abstract trusted boundary. TM3, therefore, uses the Nokia deployment to motivate the plausibility of exposure to RAN forwarding infrastructure, while the command-tampering experiment is ethically instantiated in our controlled Open5GS/UERANSIM testbed.

Separately, during our attack experiments on the Open5GS/UERANSIM testbed, we also identified and disclosed CVE-2026-7183 in UERANSIM~\cite{mitre_cve-2026-7183_2026}. The vulnerability affects UERANSIM through version 3.2.7 and allows remote crashing of the gNodeB process via an unhandled exception in the \texttt{rls::DecodeRlsMessage} function within the Radio Link Simulation Layer component when handling a malformed \texttt{pduLength} argument. The fix was incorporated in version 3.2.8. We did not use this flaw as the basis for TM3. Even so, the CVE reinforces the broader observation that software RAN components expose concrete attack surfaces, and that UAS C2 security may depend on assumptions located below the application layer.

In the 5G security architecture, primary authentication and key establishment are anchored in the UE and the core network during registration \cite{3gpp_ts_33501}. The gNodeB sits within the operator RAN trusted perimeter. For UAV C2, the consequences are significant. User Plane PDCP terminates at the gNodeB. After PDCP deciphering, the forwarder sees the application payload before encapsulation in GTP-U on N3. Absent end-to-end C2 integrity, control of that point allows the adversary to change what the autopilot will consume.

Protections on N3 vary across 5G deployments. 3GPP defines procedures for this interface, and some designs place IPSec termination at an operator-side security gateway~\cite{3gpp_ts_33501}. However, N3 transport protection and User Plane integrity remain deployment-dependent and are often disabled in commercial practice~\cite{lasierra_fact-checking_2025,noauthor_bigmac_2023}. More importantly for TM3, they do not protect against an adversary already placed after PDCP deciphering on the gNodeB. Such a forwarder operates on cleartext before GTP-U encapsulation, so we do not treat these mechanisms as substitutes for application-layer authenticity.

\begin{figure}[H]
\centering
\includegraphics[width=\columnwidth]{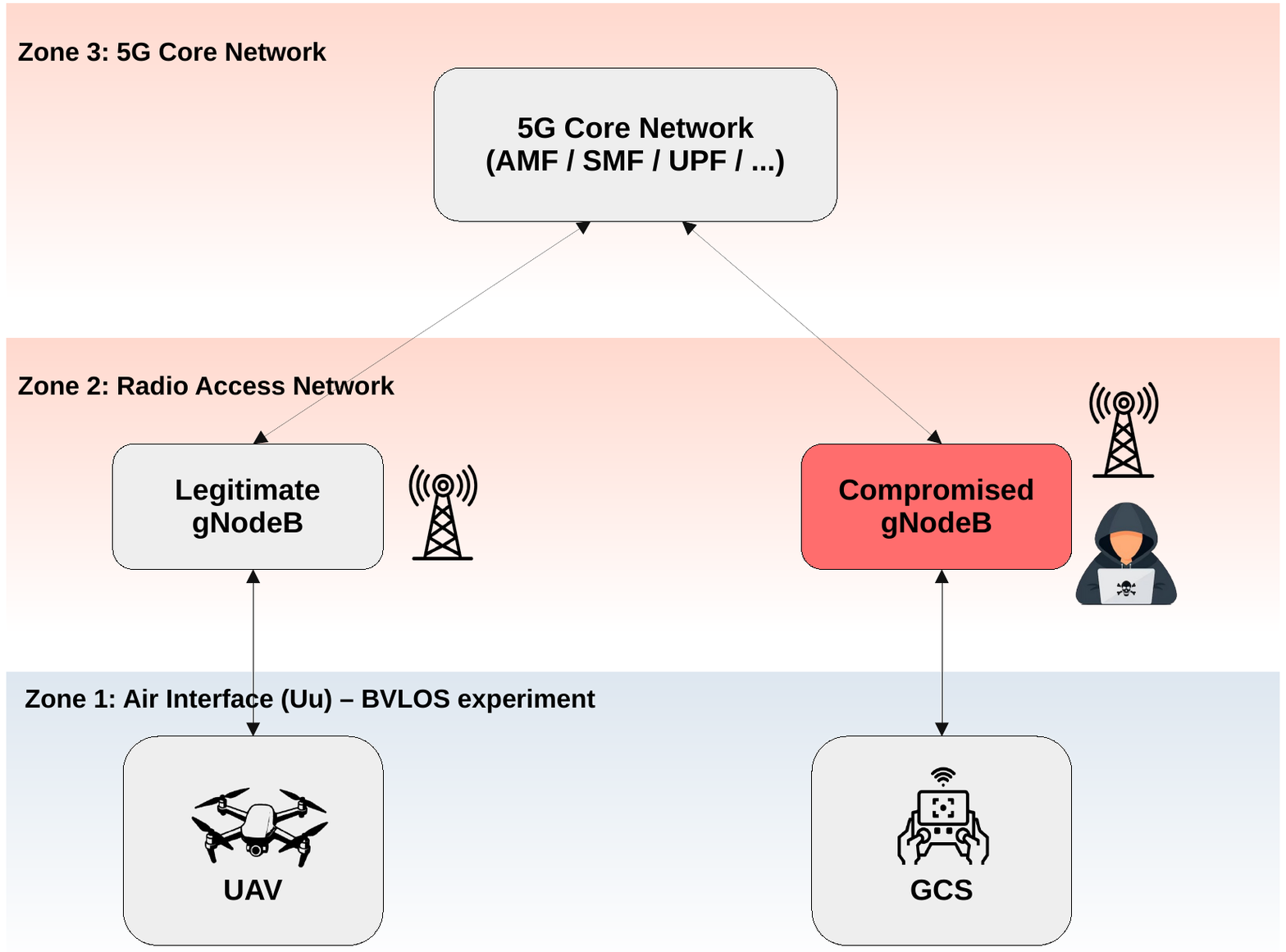}
\caption{TM3 compromised gNodeB placement in the UAV C2 path.}
\label{fig:threat-model-03}
\end{figure}

To test this configuration, we adapted our testbed (see Figure~\ref{fig:threat-model-03}) to include two gNodeBs. One is a legitimate gNodeB where the UAV is attached, and the other gNodeB is also legitimate, where the GCS is attached; however, we assume that an adversary has full access to it, that is, it is compromised. All C2 originating at the GCS crosses this compromised forwarder before N3 toward the UPF. We keep typical N3 hardening in the baseline of this testbed to reflect limitations of common 5G deployments \cite{lasierra_fact-checking_2025}.

The adversary's goal is to change the UAV's behavior without tearing down the tunnel, that is, to compromise integrity. They preserve connectivity and swap the command content. Messages such as \texttt{SET\_POSITION\_TARGET\_LOCAL\_NED} carry navigation. Without MAVLink~2 signing, there is no end-to-end authenticity of the payload \cite{ardupilot_mavlink2_signing}. With a privileged \texttt{shell} on the gNodeB, the attacker inspects and alters the inner UDP after PDCP deciphering and before re-forwarding. The GCS still sees an apparently normal control session.

With access on the compromised gNodeB, we divert GTP-U traffic on UDP/2152 to a user-space interceptor using \texttt{iptables} and \texttt{NFQUEUE}. The script parses inner MAVLink UDP, tracks \texttt{GLOBAL\_POSITION\_INT}, and rewrites \texttt{SET\_POSITION\_TARGET\_LOCAL\_NED} into a \texttt{MAV\_FRAME\_LOCAL\_OFFSET\_NED} displacement chosen by the adversary. No radio key is broken; the failure is that 5G integrity ends at the gNodeB, while command semantics must survive beyond it.

With this scenario executed in our experiment, we observed navigation hijack. The trace at the GCS looks benign. The vehicle follows the attacker's intent. The effect is semantic and stems from hop-by-hop trust \cite{lasierra_fact-checking_2025,noauthor_bigmac_2023}. The planar NED remapping uses a BVLOS-consistent approximation for the experiment scale.

These experiments show that UAS C2 can have its integrity compromised while connectivity remains stable. Between the UAV and the GCS, the MAVLink flow proceeds normally, and the GCS observes a seemingly normal control session. However, after PDCP deciphering on a compromised gNodeB, the command's semantic content can be rewritten before it reaches the autopilot. This exposes an important cross-layer gap: 5G protects network access and the radio link, yet safe UAS operation requires command integrity and authenticity that survive the forwarding infrastructure.

To test mitigations for this problem, we evaluated two mechanisms. First, we introduced IPSec on the N3 interface to protect GTP-U transport between the gNodeB and the UPF. Nevertheless, this mitigation had no effect because, as explained previously, the adversary on the compromised gNodeB can alter packets after PDCP deciphering and before encapsulation on N3. The second mitigation we tested was deploying MAVLink signing on the UAV and GCS \cite{ardupilot_mavlink2_signing}, which blocked tampered commands and accepted only commands validly signed by the GCS.

\begin{figure}[H]
\centering
\includegraphics[width=\columnwidth]{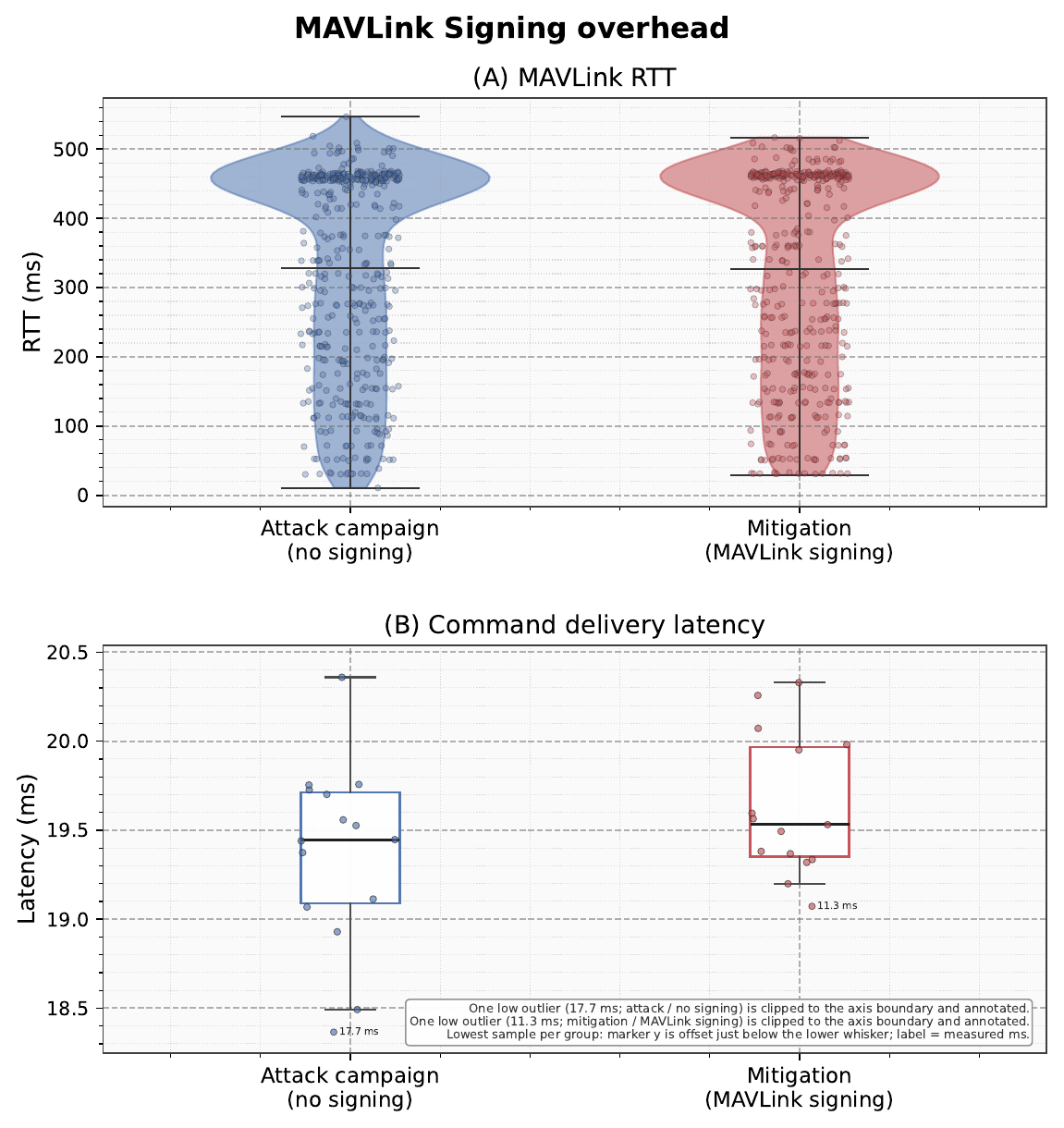}
\caption{TM3 MAVLink RTT and command latency with and without signing.}
\label{fig:threat-model-03-signing-overhead}
\end{figure}

As in TM2, we also measured the overhead introduced by MAVLink signing to verify whether enabling it causes a noticeable impact on UAS operations. We performed three five-minute runs for each condition, with and without MAVLink signing, and compared MAVLink RTT and command delivery latency over the same observation window. As shown in Figure~\ref{fig:threat-model-03-signing-overhead}, we compare the attack campaign without signing with the mitigation campaign with signing enabled. Panel~(A) shows the MAVLink RTT distribution. The two distributions remain quite similar, with overlapping central behavior and comparable spread. This indicates that MAVLink signing did not introduce a visible RTT penalty in our experimental scenario. More importantly, the mitigation changes the channel's security property without perceptibly altering its network behavior. Panel~(B) shows command delivery latency at the application layer. Latency remains concentrated around 19 to 20~ms under both conditions, with close medians and interquartile ranges. The few lower outliers were annotated because they were clipped to preserve plot readability. Thus, we did not observe perceptible latency impacts on C2. Overall, this result reinforces MAVLink signing as a practical mitigation for TM3.

MAVLink signing also introduced two practical constraints. First, it requires clock discipline. When we set the GCS clock to 01/01/2020 to emulate clock drift, the UAV rejected commands due to timestamp mismatch. Second, signing must be explicitly enabled in MAVLink~2 deployments, for example through Mission Planner or autopilot parameters~\cite{ardupilot_mavlink2_signing}. Deployments that leave signing disabled remain exposed to on-path command tampering, as shown in TM3.

To conclude, TM3 documents an integrity failure. Our experiments show that the network can remain connected, the C2 flow can continue, and the GCS can observe an apparently normal session while the UAV executes a command whose content was altered after PDCP deciphering on the compromised gNodeB. Therefore, 5G connectivity and radio-layer protection do not imply end-to-end integrity of the UAS command. Section~\ref{sec:discussion} revisits this cross-layer dependency and the need for application-layer authenticity mechanisms that survive the forwarding infrastructure.

\section{Discussion and Security Implications}
\label{sec:discussion}

This section examines the security implications of the experiments discussed earlier for BVLOS UAS in a 5G SA architecture. We do not claim that all commercial deployments fail in the same way or at the same rate. However, we present three distinct adversary scenarios along the 5G-UAV path that can violate different security guarantees while connectivity persists.

The discussion here walks through these security implications, the orthogonality with respect to 3GPP UAS association, cloud-native orchestration effects observed in experiments, a mitigation baseline, and, finally, explicit limits to generalization.

\subsection{Connectivity is not C2 UAS security}
\label{sec:discussion-connectivity}

Our experiments show that connectivity is not, by itself, a signal of operational security for BVLOS C2. Registration state and IP reachability can remain nominal while timeliness, session continuity, or semantic integrity fail.

In TM1, TM2, and TM3, the same pattern emerges across different interfaces. The 5G stack provides transport, scheduling, and session-management mechanisms. Closed-loop control treats these mechanisms as trustworthy. An adversary positioned within them can still break control without forging an association or breaking an attachment. Table~\ref{tab:tm-properties} maps each threat to the property it stresses.

\begin{table}[H]
\caption{Primary security property stressed by each threat model.}
\label{tab:tm-properties}
\centering
\small
\begin{tabular}{lc}
\hline
\textbf{Threat model} & \textbf{Primary property} \\
\hline
TM1 & Timeliness \\
TM2 & Availability \\
TM3 & Integrity \\
\hline
\end{tabular}
\end{table}

In summary, TM1 shows that timeliness can collapse under shared User Plane contention even when the UAV, the GCS, and the co-tenant adversary remain authorized on the same slice and DNN. Moreover, other variants, such as reflected attacks, may cross slice/DNN boundaries. TM2 shows that Control Plane availability during mobility can dominate whether a session survives handover long enough for MAVLink to remain usable. Finally, TM3 shows that the semantic integrity of MAVLink commands can fail with active tunnels because the trusted RAN path can still expose application octets in cleartext to a compromised forwarder. Therefore, the common failure is not the absence of connectivity but the loss of control guarantees once connectivity is established.

\subsection{UAS-specific procedures and orthogonality}
\label{sec:uas-orthogonality}

UUAA and controller-to-UAV pairing are 3GPP UAS-specific procedures that constrain which UAVs and controllers are authorized to use UAS services \cite{3gpp_ts_33256,3gpp_ts_22125,3gpp_ts_23256}. Nonetheless, they do not cover the failure modes we measured post-attachment. TM1 depends on queue contention among authorized neighbors. It does not require unauthorized association. TM2 depends on the loss of NF availability driven by an insider on the SBI surface. TM3 depends on a compromised gNodeB positioned after PDCP deciphering.

TS~33.256 places responsibility for protecting the {payload} on the USS or the UAV supplier \cite{3gpp_ts_33256}. Table~\ref{tab:uas-orthogonality} summarizes this orthogonality for the evaluated threat models:

\begin{table}[H]
\caption{Relationship between UAS procedures and the evaluated threat models.}
\label{tab:uas-orthogonality}
\centering
\small
\setlength{\tabcolsep}{4pt}
\begin{tabular*}{\columnwidth}{@{}>{\centering\arraybackslash}p{1.75cm}p{\dimexpr\columnwidth-1.75cm-2\tabcolsep\relax}@{}}
\hline
\textbf{Threat~model} & \textbf{Why UAS procedures do not prevent it} \\
\hline
TM1 & User Plane queues can saturate due to cross-traffic among authorized UAVs, controllers, and co-tenant UEs. UUAA governs the association, not per-UE contention on a shared path. \\
TM2 & An insider with SBI reachability can disrupt NRF and SMF availability. UUAA and pairing do not protect NF availability on the Control Plane. \\
TM3 & A compromised gNodeB can alter cleartext C2 payloads after PDCP deciphering unless end-to-end integrity is enforced. Access-stratum protection terminates at the gNodeB. \\
\hline
\end{tabular*}
\end{table}

When UUAA and controller-to-UAV pairing are deployed in accordance with TS~23.256 and related security procedures \cite{3gpp_ts_23256,3gpp_ts_33256}, attacks that rely on unauthorized association can be limited at the architectural level. However, our threat models do not depend on unauthorized UAS association. They target shared User Plane contention, Control Plane availability, and the trusted RAN boundary after connectivity has already been established.

\subsection{Cloud-native core platforms and orchestration semantics}
\label{sec:discussion-cloud-native}

Our TM2 runs Open5GS NFs as Kubernetes pods. For example, we observed that repeated NRF failures caused the pod to enter \texttt{CrashLoopBackOff} in TM2. Kubernetes documents this behavior as a mechanism that prolongs pod recovery time to avoid repeatedly consuming resources while the error persists \cite{kubernetes_pod_lifecycle_2026}. Although this may look testbed-specific, we argue that it is not merely an artifact of our environment. Telecom cores are increasingly discussed and deployed as containerized SBA functions on Kubernetes. This includes Kubernetes deployments on bare metal for the 5GC \cite{ericsson_kubernetes_2022}. It also includes telco cloud roadmaps that cite alignment with 3GPP, ETSI, and CNCF \cite{huawei_next-generation_2026}. Nokia also reports on 5G products running on Kubernetes across heterogeneous operator infrastructures \cite{cncf_nokia_2026}.

\subsection{Mitigation baseline}
\label{sec:mitigation-baseline}

Based on our experiments, an effective defense cannot rely on a single general solution. Each threat must be mitigated at the layer where it occurs.

For TM1, slice and DNN isolation reduce the chance that an adversarial UE shares the same bottleneck as the UAV C2 bearer. This does not eliminate contention by itself. Filtering near the UPF, rate limiting, and policing are still needed to reduce sustained cross-traffic toward the UAV or reflectors. Operator QoS differentiation, 5QI selection, GBR profiles, reflective QoS, and 5G LAN local switching are outside the scope of our quantitative study. These controls may reduce contention, but they do not, by themselves, establish that MAVLink deadlines are preserved under adversarial cross-traffic.

TM2 requires a different layer of defense. SBI segmentation, mutual TLS with strong NF identity, tight \texttt{NetworkPolicy} or equivalent ACLs, and strict separation between generic workloads and NRF or SMF paths reduce insider reachability. Robust service-state validation beyond schema checks also matters, because the failures we observed were not ordinary malformed-message rejections. Rate limiting at SBI entry points and crash-loop monitoring can further reduce the dwell time of repeated restart cycles, which our experiments showed to extend disruption windows during mobility.

For TM3, hardening transport on N3 is useful but insufficient against the threat we evaluate. A compromised gNodeB remains an endpoint that sees deciphered payloads before encapsulation. The mitigation that directly targets semantic integrity is MAVLink signing with disciplined clocks, or an equivalent application-layer authenticity mechanism.

\subsection{Limitations and scope}
\label{sec:limitations}

This subsection summarizes the main boundaries of our evaluation. We discuss the emulation model, the role of the commercial core, C2 implementation choices, mitigation depth, and assumptions that affect generalization.

\begin{itemize}
    \item \textbf{Emulation scope.} UERANSIM provides RAN and UEs without over-the-air hardware. RF effects would change numerical values. Nevertheless, they do not remove the User Plane, SBI, or PDCP boundary failure points that we exercised.

    \item \textbf{Commercial core scope.} Measurements with the Nokia core in Section~\ref{sec:threat-rogue-ue} ground TM1 reachability and traffic effects on a commercial 5G core \cite{nokia_cmu_2026}. They also support the plausibility of TM2 and TM3 by exposing deployment-relevant positions for core-side reachability and trusted RAN forwarding. We do not claim that the same TM2 robustness issues or the same TM3 forwarding path occur on that platform. Open5GS remains our reproducible substrate for controlled mechanisms of TM1, TM2, and TM3.

    \item \textbf{Closed-loop autopilot.} C2-level effects are measured through MAVLink traffic handled by Python scripts in pods. This captures timing, session continuity, and command integrity at the C2 interface. It is not, however, a full assessment with a closed-loop autopilot, such as ArduPilot SITL or HITL. We therefore report triggering \texttt{failsafe} under our C2 watchdog and treat full autopilot replication as future work.

    \item \textbf{QoS and slicing.} TM1 uses generic IP in a slice and DNN without dedicated 5QI/GBR treatment, reflective QoS, or 5G LAN local switching. We did not run a full operator QoS study. Where we applied local differentiation in the lab, it served as a controlled analog and not as a claim about normalized 5QI behavior at scale. QoS can prioritize or reserve resources for C2 traffic, but it does not, by itself, remove adversarial cross traffic or prove that MAVLink deadlines are preserved during contention.

    \item \textbf{Mitigation depth.} Functional tests show which mitigations block the attack instances we evaluated. We quantified local overhead for the deployed mitigations, including mTLS for TM2 and signing for TM3. These measurements remain testbed-scale results, not fleet-scale estimates of resource usage or operational cost.

    \item \textbf{TM3 assumptions.} We assume host-level control of a gNodeB that the core still trusts. We evaluate the impact of this position on MAVLink integrity, but not how the gNodeB is compromised or whether the compromise evades production gNodeB monitoring or O-RAN telemetry.

    \item \textbf{User Plane integrity on N3.} We did not empirically evaluate optional N3 User Plane integrity protection in the testbed. We therefore do not report performance or interoperability caveats for that mode. Even when such protection is active, it does not, by itself, restore command authenticity against a compromised gNodeB that rewrites the MAVLink payload before encapsulation. TM3, therefore, still motivates application-layer signing or an equivalent mechanism.
\end{itemize}

\section{Related Works}
\label{sec:related-works}

Prior work usually studies UAV security, 5GC robustness, GTP tunnels, rogue base stations, mobile-network queueing, MAVLink weaknesses, and security evaluation tools as separate fronts. Each line of work adopts its own success metric, such as protocol deviation, exposed interfaces, detector accuracy, or packet delay. In contrast, our work uses the UAV C2 loop as the common observation point from a cross-layer perspective.

\textbf{UAV and MAVLink security.}
Prior surveys organize threats around commercial drones, FPV links, Wi-Fi, cellular, XBee, and other control channels~\cite{nassi_sok_2021}.
Application-layer MAVLink studies examine weaknesses in authentication, timestamps, and sequence numbers~\cite{du_exploiting_2024}.
Other work shows that MAVLink can be abused to build covert channels by exploiting missing security mechanisms and redundant protocol features~\cite{veksler_catch_2024}.
These works are important for understanding direct exposure to UAVs and MAVLink.
However, they do not focus on how 5G SA mechanisms themselves can degrade the same C2 loop after the UAV and GCS are already attached to a 5G SA network.
Our work stacks MAVLink over 5G and evaluates failures caused by User Plane contention, Control Plane instability, and compromised RAN forwarding.

\textbf{5G core robustness and tunnel exposure.}
Prior work has shown that reachable GTP hosts, tunnel endpoints, and core interfaces can be abused to cause denial-of-service, session hijacking, or tunnel disruption~\cite{zhang_invade_2025,amponis_threatening_2022}.
Recent fuzzing and black-box testing of 5GC implementations also report crashes, logical deviations, and fraud-related failures~\cite{sun_5gc-fuzz_2025,dong_corecrisis_2025}.
These works mainly report core-network or protocol-level impact.
Our TM2 adds the mission-level link: NRF and SMF instability can become a UAV C2 failure when mobility requires timely session coordination.

\textbf{SBI reachability and segmentation.}
Studies of cellular infrastructure have shown that traffic from UEs or nearby network positions can sometimes cross inconsistent segmentation boundaries and reach sensitive internal elements~\cite{shaik_uncovering_2025}.
Our TM2 builds on this concern but evaluates a different consequence.
An insider with SBI reachability can affect NRF and SMF availability, and this Control Plane degradation propagates to C2 continuity during handover.

\textbf{RAN and UE attack surfaces.}
Systematic fuzzing and differential testing of base stations, handsets, and baseband behavior uncover implementation issues and security-relevant deviations~\cite{bennett_ransacked_2024,tu_logic_astray_2024}.
Other works study dishonest or rogue base stations and their impact on signaling or QoS~\cite{xu_integrity_2025,shaik_uncovering_2025}.
Complementary work proposes pre-authentication sniffing and targeted downlink injection as a third-party attacker model, explicitly contrasting with deploying a rogue base station, and demonstrates outcomes such as UE crashes, downgrade, and identity extraction~\cite{luo_sni5gect_2025}.
These efforts usually assume malformed protocol inputs, detectable radio-layer misuse, or unauthorized access nodes.
TM3 instead assumes a strategic adversary who already controls a legitimate gNodeB.
The contribution is not another fuzzing campaign but an end-to-end demonstration of how a trusted forwarding node can rewrite MAVLink semantics after PDCP deciphering while the tunnel remains active.

\textbf{Transport and RAN hardening.}
Work on disaggregated RAN and fronthaul integrity shows that software-defined RAN components enlarge the trusted computing base and can affect performance across cells~\cite{xing_criticality_2024}.
This motivates hardening of RAN transport and deployment boundaries.
Nonetheless, protecting later links does not substitute for end-to-end command authenticity once the forwarding endpoint itself is compromised.
TM3 shows that N3 or transport protection does not prevent a compromised gNodeB from modifying payloads before encapsulation.

\textbf{Timeliness and queueing in mobile networks.}
Measurements of bufferbloat and mobile-network RTT inflation show that queues can grow and produce large delay under load~\cite{jiang_tackling_2012,allman_comments_2013}.
More recent work discusses active queue management and latency control in 5G and later systems~\cite{stoltidis_aqm_2025}.
These studies explain the queueing mechanism.
Conversely, our TM1 adds an adversarial and cyber-physical reading: an authorized co-tenant UE in the same slice and DNN can weaponize shared User Plane resources against MAVLink deadlines while connectivity remains available.

\textbf{Slicing and UAS procedures.}
Prior work on 5G slicing organizes threats and open questions around intra-slice and inter-slice settings~\cite{olimid_5g_2020}.
3GPP UAS procedures address authentication, authorization, and controller-to-UAV association~\cite{3gpp_ts_23256,3gpp_ts_33256,3gpp_ts_22125}.
These mechanisms are essential, but they do not directly address the post-attachment failures studied here.
In contrast, our work shows that TM1 does not require unauthorized association, TM2 targets SBI and session availability, and TM3 targets command semantics after PDCP termination.

\textbf{Security evaluation tools and metrics.}
Replay, mutation, and anomaly-detection frameworks for 5G evaluate corrupted messages, protocol deviations, detector performance, or model precision~\cite{salazar_5greplay_2021,manca2026sage5gcsecurityawareguidelinesevaluating}.
Our work uses a different success metric.
We measure whether the UAV C2 loop suffers delayed commands, timeouts, \texttt{failsafe}, or navigation hijack while 5G connectivity indicators remain plausible.

\textbf{Cross-layer trust chains.}
Attacks originating from peripheral or apparently isolated elements, such as malicious SIM software reaching modem or baseband components, reinforce the idea that cellular systems contain long trust chains~\cite{lisowski_simurai_2024}.
Our threat models follow the same principle at different layers of 5G SA.
TM1 stresses the User Plane, TM2 stresses the Control Plane, and TM3 stresses the trusted RAN forwarding boundary.

To the best of our knowledge, prior work does not jointly evaluate, on a single 5G SA UAV C2 substrate, the triad of User Plane contention among authorized UEs, SBA-function instability under mobility, and command rewriting after PDCP termination.
The contribution of this paper is therefore not reducible to one protocol, one CVE, or one interface.
It is a cross-layer mapping showing how mechanisms that are usually studied separately can produce post-attachment C2 failure, while connectivity indicators remain misleadingly stable.

\section{Conclusion}
\label{sec:conclusion}

We demonstrated three post-attachment logical failure modes for MAVLink C2 over 5G SA under continued UE connectivity. TM1 shows that co-tenant contention can make C2 unsafe despite an active PDU Session. TM2 shows that Control Plane instability during handover can break session continuity. TM3 shows that a compromised but trusted gNodeB can rewrite MAVLink commands after PDCP deciphering, preserving the tunnel while violating command integrity.

Together, these results show that connectivity metrics alone are not evidence of UAS operational security. Safe UAV C2 requires testing closed-loop behavior under adversarial contention, insider SBI reachability, and RAN compromise, and then stacking mitigations across layers, including slice/DNN isolation, path filtering and rate limiting, SBI segmentation, mTLS with per-NF identity, robust state validation, and MAVLink signing. Connectivity is necessary for UAS operations, but our results show that it is not sufficient to demonstrate that control remains safe.

\section{Ethical Considerations}
\label{sec:ethics}

All experiments in this work were performed in controlled testbeds operated by us. We did not run attacks against public mobile networks, third-party infrastructure, production UAV systems, or equipment outside our authorization. The Nokia CMU 5G SA setup was used only to ground reachability and traffic-placement assumptions in a commercial core; we did not attempt to exploit vulnerabilities in that platform.

During the study, we found robustness issues in Open5GS and UERANSIM. These issues were reported to the maintainers before submission. Five CVEs have been assigned (CVE-2026-8187, CVE-2026-8186, CVE-2026-38511, CVE-2026-38512, and CVE-2026-7183), and the corresponding CVE records are cited. 

The artifact is meant to support reproducibility in controlled environments. We avoid including exploit-ready instructions for operational networks and frame the experiments as defensive evaluation of UAV C2 over 5G.

\bibliographystyle{IEEEtran}
\bibliography{References}

@inproceedings{nassi_sok_2021,
    author = {Nassi, Ben and Bitton, Ron and Masuoka, Ryusuke and Shabtai, Asaf and Elovici, Yuval},
    title = {{SoK}: Security and Privacy in the Age of Commercial Drones},
    booktitle = {2021 {IEEE} Symposium on Security and Privacy ({SP})},
    year = {2021},
    month = {May},
    pages = {1434--1451},
    doi = {10.1109/SP40001.2021.00005},
}

@inproceedings{xing_criticality_2024,
    author = {Xing, Jiarong and Yoo, Sophia and Foukas, Xenofon and Kim, Daehyeok and Reiter, Michael K.},
    title = {On the Criticality of Integrity Protection in {5G} Fronthaul Networks},
    booktitle = {33rd {USENIX} Security Symposium ({USENIX} Security 24)},
    year = {2024},
    pages = {4463--4479},
    url = {https://www.usenix.org/conference/usenixsecurity24/presentation/xing-jiarong}
}

@inproceedings{noauthor_bigmac_2023,
    author = {Heijligenberg, Thijs and Knips, Guido and B\"{o}hm, Christian and Rupprecht, David and Kohls, Katharina},
    title = {{BigMac}: Performance Overhead of User Plane Integrity Protection in {5G} Networks},
    booktitle = {Proceedings of the 16th {ACM} Conference on Security and Privacy in Wireless and Mobile Networks},
    year = {2023},
    pages = {145--150},
    doi = {10.1145/3558482.3581777},
}

@misc{xu_integrity_2025,
    author = {Xu, Jiali and Loscri, Valeria and Rouvoy, Romain},
    title = {Integrity Under Siege: A Rogue {gNodeB}'s Manipulation of {5G} Network Slice Allocation},
    year = {2025},
    howpublished = {arXiv:2511.03312},
    url = {https://arxiv.org/abs/2511.03312},
    note = {arXiv preprint}
}

@inproceedings{luo_sni5gect_2025,
    author = {Luo, Shijie and Garbelini, Matheus and Chattopadhyay, Sudipta and Zhou, Jianying},
    title = {{SNI5GECT}: A Practical Approach to Inject {aNRchy} into {5G} {NR}},
    booktitle = {34th {USENIX} Security Symposium ({USENIX} Security 25)},
    year = {2025},
    pages = {5385--5404},
    url = {https://www.usenix.org/conference/usenixsecurity25/presentation/luo-shijie}
}

@inproceedings{lisowski_simurai_2024,
    author = {Lisowski, Tomasz Piotr and Chlosta, Merlin and Wang, Jinjin and Muench, Marius},
    title = {{SIMurai}: Slicing Through the Complexity of {SIM} Card Security Research},
    booktitle = {33rd {USENIX} Security Symposium ({USENIX} Security 24)},
    year = {2024},
    pages = {4481--4498},
    url = {https://www.usenix.org/conference/usenixsecurity24/presentation/lisowski}
}

@inproceedings{dong_corecrisis_2025,
    author = {Dong, Yilu and Yang, Tianchang and Ishtiaq, Abdullah Al and Rashid, Syed Md Mukit and Ranjbar, Ali and Tu, Kai and Wu, Tianwei and Mahmud, Md Sultan and Hussain, Syed Rafiul},
    title = {{CoreCrisis}: Threat-Guided and Context-Aware Iterative Learning and Fuzzing of {5G} Core Networks},
    booktitle = {34th {USENIX} Security Symposium ({USENIX} Security 25)},
    year = {2025},
    pages = {5287--5306},
    url = {https://www.usenix.org/conference/usenixsecurity25/presentation/dong-yilu}
}

@inproceedings{veksler_catch_2024,
    author = {Veksler, Maryna and Akkaya, Kemal and Uluagac, Selcuk},
    title = {Catch me if you can: Covert Information Leakage from Drones using {MAVLink} Protocol},
    booktitle = {Proceedings of the 19th {ACM} Asia Conference on Computer and Communications Security},
    publisher = {Association for Computing Machinery},
    address = {New York, NY, USA},
    year = {2024},
    month = {Jul},
    pages = {902--914},
    doi = {10.1145/3634737.3637672},
}

@techreport{3gpp_ts_33501,
    author = {{3GPP}},
    title = {{TS 33.501: Security architecture and procedures for 5G System (Release 17)}},
    institution = {3rd Generation Partnership Project (3GPP)},
    year = {2022}
}

@techreport{3gpp_ts_33256,
    author = {{3GPP}},
    title = {{TS 33.256: Security aspects of Uncrewed Aerial Systems (UAS) (Release 17)}},
    year = {2022},
    institution = {3rd Generation Partnership Project (3GPP)}
}

@techreport{3gpp_ts_29244,
    author = {{3GPP}},
    title = {{TS 29.244: Interface between the Control Plane and the User Plane nodes (PFCP) (Release 17)}},
    institution = {3rd Generation Partnership Project (3GPP)},
    year = {2022}
}

@techreport{3gpp_ts_23502,
    author = {{3GPP}},
    title = {{TS 23.502: Procedures for the 5G System (5GS) (Release 17)}},
    institution = {3rd Generation Partnership Project (3GPP)},
    year = {2022}
}

@techreport{3gpp_ts_23501,
    author = {{3GPP}},
    title = {{TS 23.501: System architecture for the 5G System (5GS) (Release 17)}},
    institution = {3rd Generation Partnership Project (3GPP)},
    year = {2022}
}

@techreport{3gpp_ts_23256,
    author = {{3GPP}},
    title = {{TS 23.256: Support of Uncrewed Aerial Systems (UAS) connectivity, identification and tracking; Stage 2 (Release 17)}},
    institution = {3rd Generation Partnership Project (3GPP)},
    year = {2022}
}

@techreport{3gpp_ts_22261,
    author = {{3GPP}},
    title = {{TS 22.261: Service requirements for the 5G system (Release 17)}},
    year = {2017},
    institution = {3rd Generation Partnership Project (3GPP)}
}

@techreport{3gpp_ts_22125,
    author = {{3GPP}},
    title = {{TS 22.125: Unmanned Aerial System (UAS) support in 3GPP (Release 17)}},
    year = {2022},
    institution = {3rd Generation Partnership Project (3GPP)}
}

@techreport{3gpp_tr_33891,
    author = {{3GPP}},
    title = {{TR 33.891: Study on security of phase 2 for UAS, UAV and UAM (Release 18)}},
    year = {2023},
    institution = {3rd Generation Partnership Project (3GPP)}
}

@techreport{3gpp_tr_33809,
    author = {{3GPP}},
    title = {{TR 33.809: Study on 5G security enhancements against False Base Stations (FBS) (Release 17)}},
    year = {2023},
    institution = {3rd Generation Partnership Project (3GPP)}
}

@inproceedings{sun_5gc-fuzz_2025,
    author = {Sun, Yu and Liu, Xinyu and Sun, Qian and Wang, Jiaming and Tian, Lin and Liu, Jianwei},
    title = {{5GC-Fuzz}: Finding Deep Stateful Vulnerabilities in {5G} Core Network with Black-Box Fuzzing},
    booktitle = {{IEEE} {INFOCOM} 2025 - {IEEE} Conference on Computer Communications},
    year = {2025},
    month = {May},
    pages = {1--10},
    doi = {10.1109/INFOCOM55648.2025.11044489},
}

@inproceedings{shaik_uncovering_2025,
    author = {Shaik, Altaf and Jaschek, Robert and Seifert, Jean-Pierre},
    title = {Uncovering Hidden Paths in {5G}: Exploiting Protocol Tunneling and Network Boundary Bridging},
    booktitle = {Proceedings of the 2025 {ACM} {SIGSAC} Conference on Computer and Communications Security},
    publisher = {Association for Computing Machinery},
    address = {New York, NY, USA},
    year = {2025},
    month = {Nov},
    pages = {231--245},
    doi = {10.1145/3719027.3765206},
}

@inproceedings{zhang_invade_2025,
    author = {Zhang, Yiming and Wan, Tao and Yang, Yaru and Duan, Haixin and Wang, Yichen and Chen, Jianjun and Wei, Zixiang and Li, Xiang},
    title = {Invade the Walled Garden: Evaluating {GTP} Security in Cellular Networks},
    booktitle = {2025 {IEEE} Symposium on Security and Privacy ({SP})},
    year = {2025},
    month = {May},
    pages = {1159--1177},
    doi = {10.1109/SP61157.2025.00028},
}

@inproceedings{singh_overview_2022,
    author = {Singh, Radheshyam and Ballal, Kalpit Dilip and Berger, Michael Stübert and Dittmann, Lars},
    title = {Overview of Drone Communication Requirements in {5G}},
    booktitle = {Internet of Things},
    publisher = {Springer International Publishing},
    address = {Cham},
    year = {2022},
    pages = {3--16},
    doi = {10.1007/978-3-031-20936-9_1}
}

@article{olimid_5g_2020,
    author = {Olimid, Ruxandra F. and Nencioni, Gianfranco},
    title = {{5G} Network Slicing: A Security Overview},
    journal = {{IEEE} Access},
    volume = {8},
    pages = {99999--100009},
    year = {2020},
    doi = {10.1109/ACCESS.2020.2997702},
}

@article{amponis_threatening_2022,
    author = {Amponis, George and Radoglou-Grammatikis, Panagiotis and Lagkas, Thomas and Mallouli, Wissam and Cavalli, Ana and Klonidis, Dimitris and Markakis, Evangelos and Sarigiannidis, Panagiotis},
    title = {Threatening the {5G} Core via {PFCP} {DoS} Attacks: The Case of Blocking {UAV} Communications},
    journal = {EURASIP Journal on Wireless Communications and Networking},
    volume = {2022},
    number = {1},
    pages = {124},
    year = {2022},
    month = {Dec},
    doi = {10.1186/s13638-022-02204-5},
}

@inproceedings{salazar_5greplay_2021,
    author = {Salazar, Zujany and Nguyen, Huu Nghia and Mallouli, Wissam and Cavalli, Ana R. and Montes de Oca, Edgardo},
    title = {{5Greplay}: A {5G} Network Traffic Fuzzer - Application to Attack Injection},
    booktitle = {Proceedings of the 16th International Conference on Availability, Reliability and Security},
    publisher = {Association for Computing Machinery},
    address = {New York, NY, USA},
    year = {2021},
    month = {Aug},
    pages = {1--8},
    doi = {10.1145/3465481.3470079},
}

@inproceedings{du_exploiting_2024,
    author = {Du, Fei and Ge, Jinai and Wang, Wen and Zou, Yuwen and Chang, Sang-Yoon and Fan, Wenjun},
    title = {Exploiting the Vulnerabilities in {MAVLink} Protocol for {UAV} Hijacking},
    booktitle = {2024 17th International Conference on Security of Information and Networks ({SIN})},
    year = {2024},
    month = {Dec},
    pages = {1--8},
    doi = {10.1109/SIN63213.2024.10871546},
}

@article{branco_cyber_2025,
    author = {Branco, Bruno and Silvestre Serra Silva, José and Correia, Miguel},
    title = {Cyber Attacks on Commercial Drones: A Review},
    journal = {{IEEE} Access},
    volume = {13},
    pages = {9566--9577},
    year = {2025},
    doi = {10.1109/ACCESS.2025.3527698},
}

@article{geraci_what_2022,
    author = {Geraci, Giovanni and Garcia-Rodriguez, Adrian and Azari, M. Mahdi and Lozano, Angel and Mezzavilla, Marco and Chatzinotas, Symeon and Chen, Yun and Rangan, Sundeep and Renzo, Marco Di},
    title = {What Will the Future of {UAV} Cellular Communications Be? A Flight From {5G} to {6G}},
    journal = {{IEEE} Communications Surveys \& Tutorials},
    volume = {24},
    number = {3},
    pages = {1304--1335},
    year = {2022},
    doi = {10.1109/COMST.2022.3171135},
}

@article{lasierra_fact-checking_2025,
    author = {Lasierra, Oscar and Ludant, Norbert and Garcia-Aviles, Gines and Municio, Esteban and Noubir, Guevara and Skarmeta, Antonio and Costa-P{\'e}rez, Xavier},
    title = {Fact-Checking {5G} Security: Bridging the Gap Between Expectations and Reality},
    journal = {{IEEE} Open Journal of the Communications Society},
    volume = {6},
    pages = {6242--6257},
    year = {2025},
    doi = {10.1109/OJCOMS.2025.3593140},
}

@inproceedings{giambartolomei_penetration_2024,
    author = {Giambartolomei, Filippo and Barcel{\'o}, Marc and Brighente, Alessandro and Urbieta, Aitor and Conti, Mauro},
    title = {Penetration Testing of {5G} Core Network Web Technologies},
    booktitle = {{ICC} 2024 - {IEEE} International Conference on Communications},
    year = {2024},
    month = {Jun},
    pages = {702--707},
    doi = {10.1109/ICC51166.2024.10622336},
}

@inproceedings{bennett_ransacked_2024,
    author = {Bennett, Nathaniel and Zhu, Weidong and Simon, Benjamin and Kennedy, Ryon and Enck, William and Traynor, Patrick and Butler, Kevin R. B.},
    title = {{RANsacked}: A Domain-Informed Approach for Fuzzing {LTE} and {5G} {RAN}-Core Interfaces},
    booktitle = {Proceedings of the 2024 on {ACM} {SIGSAC} Conference on Computer and Communications Security},
    publisher = {Association for Computing Machinery},
    address = {New York, NY, USA},
    year = {2024},
    month = {Dec},
    pages = {2027--2041},
    doi = {10.1145/3658644.3670320},
}

@article{baguer_enabling_2024,
    author = {Baguer, Pau and Municio, Esteban and Garcia-Aviles, Gines and Costa-Pérez, Xavier},
    title = {Enabling Beyond-Visual-Line-of-Sight Drones Operation Over Open {RAN} {5G} Networks With Slicing},
    journal = {{IEEE} Network},
    volume = {38},
    number = {6},
    pages = {163--169},
    year = {2024},
    month = {Nov},
    doi = {10.1109/MNET.2024.3420085},
}

@misc{mavlink_guide,
    author = {{MAVLink Development Team}},
    title = {{MAVLink} Guide},
    year = {2025},
    url = {https://mavlink.io/},
    note = {Accessed: 2026-05-06}
}

@misc{open5gs,
    author = {{Open5GS}},
    title = {{Open5GS}: An Open Source C-language Implementation of {5G} Core and {EPC}},
    year = {2026},
    url = {https://open5gs.org/},
    note = {Accessed: 2026-05-06}
}

@misc{kubernetes,
    author = {{Cloud Native Computing Foundation}},
    title = {Kubernetes},
    year = {2026},
    url = {https://kubernetes.io/},
    note = {Accessed: 2026-05-06}
}

@misc{kind,
    author = {{The Kubernetes Authors}},
    title = {kind: Kubernetes {IN} Docker},
    year = {2026},
    url = {https://kind.sigs.k8s.io/},
    note = {Accessed: 2026-05-06}
}

@misc{docker,
    author = {{Docker Inc.}},
    title = {Docker},
    year = {2026},
    url = {https://www.docker.com/},
    note = {Accessed: 2026-05-06}
}

@misc{open5gs_operator,
    author = {{Gradiant}},
    title = {{Open5GS} Operator for Kubernetes},
    year = {2026},
    url = {https://gradiant.github.io/open5gs-operator/},
    note = {Accessed: 2026-05-06}
}

@misc{ueransim,
    author = {Gungor, Ali},
    title = {{UERANSIM}: Open Source {5G} {UE} and {RAN} ({gNodeB}) Simulator},
    year = {2026},
    url = {https://github.com/aligungr/UERANSIM},
    note = {Accessed: 2026-05-06}
}

@misc{pymavlink,
    author = {{ArduPilot}},
    title = {pymavlink: Python {MAVLink} Interface and Utilities},
    year = {2026},
    url = {https://github.com/ArduPilot/pymavlink},
    note = {Accessed: 2026-05-06}
}

@misc{ardupilot_mavlink2_signing,
    author = {{ArduPilot Dev Team}},
    title = {{MAVLink2} Signing},
    year = {2024},
    url = {https://ardupilot.org/sub/docs/common-MAVLink2-signing.html},
    note = {Accessed: 2026-05-06}
}

@misc{px4_mavlink_encryption_2025,
    author = {{PX4 Autopilot Community}},
    title = {{MAVLink} Encryption},
    howpublished = {PX4 Discussion Forum thread (opened by user Ajeet1)},
    year = {2025},
    month = {Dec},
    url = {https://discuss.px4.io/t/mavlink-encryption/48170},
    note = {Online forum, accessed: 2026-05-06}
}

@misc{manca2026sage5gcsecurityawareguidelinesevaluating,
    author = {Manca, Cristian and Scano, Christian and Piras, Giorgio and Brau, Fabio and Pintor, Maura and Biggio, Battista},
    title = {{SAGE-5GC}: Security-Aware Guidelines for Evaluating Anomaly Detection in the {5G} Core Network},
    year = {2026},
    howpublished = {arXiv:2602.03596},
    url = {https://arxiv.org/abs/2602.03596},
    note = {arXiv preprint}
}

@article{allman_comments_2013,
    author = {Allman, Mark},
    title = {Comments on Bufferbloat},
    journal = {{ACM} {SIGCOMM} Computer Communication Review},
    volume = {43},
    number = {1},
    pages = {30--37},
    year = {2013},
    month = {Jan},
    doi = {10.1145/2427036.2427041},
}

@inproceedings{jiang_tackling_2012,
    author = {Jiang, Haiqing and Wang, Yaogong and Lee, Kyunghan and Rhee, Injong},
    title = {Tackling Bufferbloat in {3G}/{4G} Networks},
    booktitle = {Proceedings of the 2012 {ACM} Conference on Internet Measurement Conference ({IMC} '12)},
    publisher = {Association for Computing Machinery},
    address = {New York, NY, USA},
    year = {2012},
    month = {Nov},
    pages = {329--342},
    doi = {10.1145/2398776.2398810},
}

@article{stoltidis_aqm_2025,
    author = {Stoltidis, Alexandros and Choumas, Kostas and Korakis, Thanasis},
    title = {Active queue management in {5G} and beyond cellular networks using Machine Learning},
    journal = {Computer Communications},
    volume = {236},
    year = {2025},
    pages = {108108},
    doi = {10.1016/j.comcom.2025.108108}
}

@misc{kubernetes_pod_lifecycle_2026,
    author = {{Kubernetes}},
    title = {Pod Lifecycle},
    year = {2026},
    url = {https://kubernetes.io/docs/concepts/workloads/pods/pod-lifecycle/},
    note = {Accessed: 2026-05-06},
}

@inproceedings{tu_logic_astray_2024,
    author = {Tu, Kai and Al Ishtiaq, Abdullah and Rashid, Syed Md Mukit and Dong, Yilu and Wang, Weixuan and Wu, Tianwei and Hussain, Syed Rafiul},
    title = {Logic Gone Astray: A Security Analysis Framework for the Control Plane Protocols of {5G} Basebands},
    booktitle = {33rd {USENIX} Security Symposium ({USENIX} Security 24)},
    year = {2024},
    month = {Aug},
    pages = {3063--3080},
    url = {https://www.usenix.org/conference/usenixsecurity24/presentation/tu}
}

@misc{nokia_cmu_2026,
    author = {{Nokia}},
    title = {Compact Mobility Unit ({CMU})},
    year = {2026},
    url = {https://www.dac.nokia.com/connectivity-solutions/compact-mobility-unit-cmu/},
    note = {Accessed: 2026-05-06}
}

@misc{ericsson_kubernetes_2022,
    author = {{Ericsson}},
    title = {Kubernetes over bare metal cloud infrastructure -- why it's important and what you need to know},
    year = {2022},
    month = {May},
    url = {https://www.ericsson.com/en/blog/2022/5/kubernetes-over-bare-metal-cloud-infrastructure-why-its-important-and-what-you-need-to-know},
    note = {Accessed: 2026-05-06}
}

@misc{huawei_next-generation_2026,
    author = {{Huawei}},
    title = {Next-Generation Telco Cloud Based on Cloud Native},
    year = {2026},
    url = {https://carrier.huawei.com/en/industry-perspective/5g-core-network/next-generation-telco-cloud-based-on-cloud-native},
    note = {Accessed: 2026-05-06}
}

@misc{cncf_nokia_2026,
    author = {{Cloud Native Computing Foundation}},
    title = {Nokia Case Study},
    year = {2026},
    url = {https://kubernetes.io/case-studies/nokia/},
    note = {Accessed: 2026-05-06}
}

@misc{huawei_ar502h-5g_2026,
    author = {{Huawei}},
    title = {{AR502H-5G} Router / {IoT} Gateway - Support},
    year = {2026},
    url = {https://support.huawei.com/enterprise/en/routers/ar502h-5g-pid-251982821},
    note = {Accessed: 2026-05-06}
}

@misc{mitre_cve-2026-7183_2026,
    author = {{The MITRE Corporation}},
    title = {{CVE-2026-7183}: aligungr {UERANSIM} Radio Link Simulation Layer rls\_pdu.cpp DecodeRlsMessage uncaught exception},
    year = {2026},
    month = {Apr},
    url = {https://www.cve.org/CVERecord?id=CVE-2026-7183}
}

@misc{mitre_cve-2026-8186_2026,
    author = {{The MITRE Corporation}},
    title = {{CVE-2026-8186}: {Open5GS} {UPF} {GTP-U} data-plane issues leading to remote denial of service ({SIGSEGV})},
    year = {2026},
    month = {May},
    url = {https://www.cve.org/CVERecord?id=CVE-2026-8186}
}

@misc{mitre_cve-2026-8187_2026,
    author = {{The MITRE Corporation}},
    title = {{CVE-2026-8187}: {Open5GS} {UPF} severe user-plane degradation via crafted high-rate {GTP-U} traffic},
    year = {2026},
    month = {May},
    url = {https://www.cve.org/CVERecord?id=CVE-2026-8187}
}

@misc{mitre_cve-2026-38511_2026,
    author = {{The MITRE Corporation}},
    title = {{CVE-2026-38511}: {Open5GS} {NRF} reachable assertion in ogs\_sbi\_getpath\_from\_uri leading to remote denial of service},
    year = {2026},
    month = {Jul},
    url = {https://www.cve.org/CVERecord?id=CVE-2026-38511}
}

@misc{mitre_cve-2026-38512_2026,
    author = {{The MITRE Corporation}},
    title = {{CVE-2026-38512}: {Open5GS} {SMF} reachable assertion in smf\_gsm\_state\_operational leading to remote denial of service},
    year = {2026},
    month = {Jul},
    url = {https://www.cve.org/CVERecord?id=CVE-2026-38512}
}

\end{document}